\documentclass[aps,superscriptaddress,twocolumn,floatfix,longbibliography]{revtex4-1}
\usepackage{dcolumn}
\usepackage{graphicx,color}
\usepackage[utf8]{inputenc}
\usepackage{amsmath,amssymb,bm,amsfonts,dsfont,mathrsfs,amsthm}
\usepackage{braket}
\usepackage{txfonts,comment}
\usepackage[colorlinks=true,linkcolor=blue,citecolor=blue,urlcolor=blue]{hyperref}
\usepackage{soul}

\begin{document}

\title{Scalable Phonon-Laser Arrays with Self-Organized Synchronization}

\author{Hugo Molinares}
\email{hugo.molinares@ufrontera.cl}
\affiliation{Departamento de Ciencias F\'{\i}sicas, Universidad de La Frontera, Casilla 54-D, Temuco, Chile}

\author{Guillermo Romero}
\email{guillermo.romero@usach.cl}
\affiliation{Departamento de F\'{\i}sica, CEDENNA, Universidad de Santiago de Chile, Avenida V\'{\i}ctor Jara 3493, 9170124, Santiago, Chile}

\author{Victor Montenegro}
\email{victor.montenegro@ku.ac.ae}
\affiliation{College of Computing and Mathematical Sciences, Department of Applied Mathematics and Sciences, Khalifa University of Science and Technology, 127788 Abu Dhabi, United Arab Emirates}
\affiliation{Institute of Fundamental and Frontier Sciences, University of Electronic Science and Technology of China, Chengdu 611731, China}
\affiliation{Key Laboratory of Quantum Physics and Photonic Quantum Information, Ministry of Education, University of Electronic Science and Technology of China, Chengdu 611731, China}

\author{Vitalie Eremeev}  
\email{vitalie.eremeev@gmail.com}
\affiliation{Centro Multidisciplinario de F\'{\i}sica, Vicerrector\'{\i}a de Investigación, Universidad Mayor, 8580745 Santiago, Chile}

\begin{abstract}
Quantum mechanical oscillators operating at frequencies up to the GHz regime have been predicted to support phonon lasing---self-sustained coherent vibrational motion emerging when the effective gain exceeds intrinsic losses. Current phonon-laser proposals face two key limitations, namely: they lack scalability and rely on coupling all oscillators to a common field, which significantly restricts flexibility and prevents selective, on-demand phonon lasing at specific locations. Given that numerous applications and theoretical insights naturally emerge from scalable many-body systems, addressing these limitations is timely. In this Letter, we demonstrate how scalable arrays of individually addressable phonon lasers can be generated through local driving in a quantum many-body Ising-like spin chain. We rigorously establish the resonance conditions under which mechanical oscillators transition from thermal motion to sustained coherent self-oscillation. Unlike previous approaches that rely on a common coupling bus, our proposal employs purely local driving, resulting in an inherently modular and scalable architecture ideally suited for integration into large-scale quantum systems. Additionally, our approach enables on-demand lasing of individual mechanical oscillators at specific sites by simply switching the spin-mechanical coupling interaction on and off, provided specific resonance conditions are satisfied. Notably, our phonon laser array is robust against resonance mismatches and naturally exhibits both pairwise self-organized synchronization and global phase locking near resonance. Finally, we outline an experimental implementation within current experimental capabilities.
\end{abstract}

\maketitle

\section{Introduction} 
Phonon lasing, the sudden transition from thermal fluctuations to the coherent stimulated emission of center-of-mass phonons, provides valuable quantum coherent states for quantum acoustics~\cite{chu2017quantum}, high-precision sensing~\cite{caves1980onthemeasurement, pan2023phonon, burd2019quantum, liu2021phonon}, and fundamental physics~\cite{bose2025massive, covey2025probing}. Phonon lasing phenomena have been extensively explored through numerous theoretical studies~\cite{wallentowitz1996vaser, lozovik2000phonon, bargatin2003nanomechanical, chudnovsky2004phonon, mendonca2010phonon, li2025dissipation, jiang2018nonreciprocal, zhang2018phase, wang2018phonon, lee2023prototype, vashahri2021magnomechanical, Eremeev2024AQT, Eremeev2020PRA, wang2018polarization, wendt2025electricallyinjectedsolidstate, lei2023quantum, khaetskii2013proposal, xie2013pulsed} and demonstrated in a wide range of experimental architectures~\cite{bron1978stimulated, liu2003coupled, vahala2009phonon, mahboob2013phonon, zhang2018phonon, ohtani2019electrically, pettit2019optical, guo2021optical, mercade2021floquet, chafatinos2020polariton, zhang2022dissipative, behrle2023phonon, ng2023intermodulation, wang2023laser2, xiao2023nonlinear, zhang2012synchronization, grudinin2010phonon, fu2023photonphonon, kuang2023nonlinear, xiong2023phonon, pan2024closed, knunz2010injection}. In these systems, similarly to conventional photonic lasing, once a well-defined threshold is crossed, the mechanical mode transitions into self-sustained coherent oscillation. Consequently, its amplitude is amplified either through stimulated phonon emission~\cite{vahala2009phonon} or optical dynamical backaction~\cite{kippenberg2005analysis,poot2012backaction}. The resulting steady-state motion is characterized by a stable limit cycle, whose phase either undergoes diffusion due to intrinsic noise~\cite{fong2014phase} or can be stabilized by a weak external coherent drive~\cite{bekker2017injection}. 

Despite significant theoretical and experimental progress, phonon-laser generation and synchronization remain restricted to a small number of oscillators coupled through a common field~\cite{zhang2012synchronization, santos2017light, bagheri2013photonic, sheng2020selforganized, ortiz2025dualmode, colombano2019synchronization, mercade2021floquet}. The above common-field approach, however, forces all oscillators to behave collectively, that is either all oscillators simultaneously exhibit lasing or none do. Consequently, one cannot selectively choose which individual oscillators exhibit lasing features, significantly reducing the flexibility of the lasing scheme. Therefore, generating arrays of phonon lasers beyond a few coupled oscillators toward scalable arrays of individually addressable phonon lasers is particularly timely. Indeed, arrays of phonon lasers could significantly advance quantum technological applications~\cite{pan2023phonon, burd2019quantum}, enable fundamental tests of quantum mechanics~\cite{bose2025massive}, facilitate long-distance synchronization protocols~\cite{li2022alloptical}, and promote exploration of quantum-acoustic many-body phenomena~\cite{chu2017quantum}. Motivated by these possibilities, we address two essential questions: (i) whether scalable and on-demand phonon-laser arrays can be generated through local driving and without coupling all oscillators to a common field; and (ii) whether such arrays remain robust against deviations in the parameters required for lasing. 

In this work, we provide affirmative answers to both questions. Indeed, we demonstrate how to generate scalable arrays of individually addressable phonon lasers through local driving in a quantum many-body Ising-like spin chain. We rigorously establish the precise resonance conditions required for phonon lasing by explicitly driving resonant blue-sideband transitions between pairs of spins and a single MO. Our findings reveal that these phonon lasers can exhibit either phase diffusion or robust phase locking in the dissipative steady-state regime. Furthermore, we demonstrate that even under conditions of resonance mismatch, our system spontaneously achieves self-organized pairwise synchronization and global phase locking~\cite{kuramoto1975self, acebron2005kuramoto}. Notably, although strongly correlated many-body systems have been extensively explored in various contexts~\cite{sachdev1999quantum, bose2003quantum, huang2020superconducting, georgescu2014quantum, montenegro2025review}, their potential for realizing multiple phonon lasers has remained largely unexplored. Hence, our results pave the way toward scalable arrays of phonon lasers utilizing strongly correlated quantum systems with time-dependent spin-exchange interactions.

\section{The model} 
\begin{figure}[t]
\centering
\includegraphics[width=\linewidth]{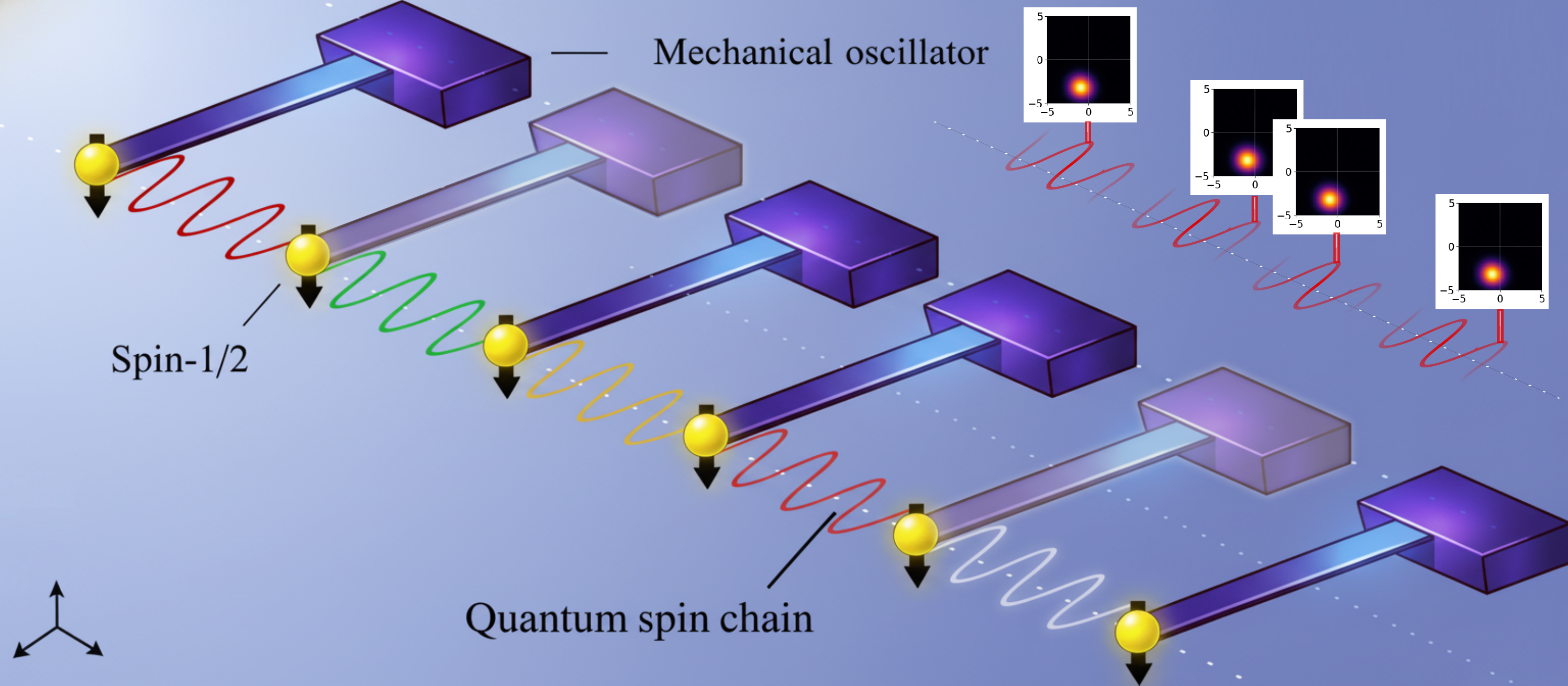}
\caption{Mechanical oscillators (MO) individually coupled to sites of a spin chain of length $N$. Local time-dependent nearest-neighbour exchange couplings at precise resonance conditions enable steady-state phonon-laser arrays. Our lasing scheme allows us to easily decouple MOs at any site (shown as semi-transparent oscillators in the sketch), thereby enabling the generation of arbitrary on-demand arrays of steady-state phonon lasers.}
\label{fig_schematics}
\end{figure}

We consider a locally driven quantum many-body Ising-like spin chain where each site is coupled to a MO, see Fig.~\ref{fig_schematics}. The Hamiltonian is $(\hbar=1)$:
\begin{eqnarray}
\hat{\mathcal{H}}&=&\sum_{j=1}^N \frac{\Delta_j}{2}\hat{\sigma}_{j}^{z}{+}\omega_j\hat{b}_j^{\dagger}\hat{b}_j{-}\lambda_{j}\hat{\sigma}_{j}^{z}(\hat{b}_j^{\dagger}{+}\hat{b}_j)\nonumber\\
&&+\sum_{j=1}^{N-1}J_{j} \cos{(\Omega_jt)} \hat{\sigma}_{j}^{x}\hat{\sigma}_{j+1}^{x},\label{eq_driven_hamiltonian}
\end{eqnarray}
where $\hat{\sigma}_j^\alpha$ ($\alpha=x,y,z$) denotes the Pauli operator acting on the spin at site $j$ along the $\alpha$ direction, and $\hat{b}_j$ ($\hat{b}^\dagger_j$) denotes the annihilation (creation) operator of the MO at site $j$. The parameter $\Delta_j$ specifies the energy splitting of the spin system, $\omega_j$ is the resonance frequency of the MO, $J_j$ is the amplitude of the external driving field, $\Omega_j$ is its driving frequency, and $\lambda_j$ quantifies the coupling strength between the spin and the MO. We consider $\Delta_j<\omega_j$ throughout. Two important points regarding the interactions in Eq.~\eqref{eq_driven_hamiltonian} must be clarified. First, the spin-boson interaction terms in Eq.~\eqref{eq_driven_hamiltonian} arise because the spin energy splitting is highly sensitive to local electric and magnetic fields. Consequently, spin-boson interactions of the form $\propto \hat{\sigma}_j^z(\hat{b}_j^\dagger{+}\hat{b}_j)$ can be engineered by integrating vibrating electrodes or magnetic tips into MOs~\cite{treutlein2014hybrid}. Second, the nature of the driven spin-spin interaction $J_{j} \cos{(\Omega_jt)} \hat{\sigma}_{j}^{x}\hat{\sigma}_{j+1}^{x}$ in Eq.~\eqref{eq_driven_hamiltonian} allows us to select resonance conditions for the simultaneous excitation (de-excitation) of spins, which represents the main mechanism for phonon lasing discussed in our work. Recently, this atypical or kinetic spin-spin coupling with zero time-average has been discussed in the context of strongly correlated bosonic lattices~\cite{pieplow2018generation,pieplow2019protected,pena2022fractional,pena2022stable,mateos2023superfluidity}, and spin systems for two-tone Floquet engineering \cite{pena2025steering}.

Lasing can only emerge when both gain and losses are present. To capture this requirement, we model the open quantum dynamics using the quantum master equation~\cite{petruccione2002theory}
\begin{eqnarray}\label{eq:master_equation}
\frac{d\hat{\rho}}{dt}&=&-i[\hat{\mathcal{H}},\hat{\rho}]+\sum_{j=1}^N\Gamma_{j}\Bigg[\left(1+\bar{n}_{j}^{s}\right)\mathcal{L}_{\hat{\sigma}_{j}^{-}}[\hat{\rho}]+\bar{n}_{j}^{s}\,\mathcal{L}_{\hat{\sigma}_{j}^{+}}[\hat{\rho}]\Bigg]
\nonumber\\
&+&\sum_{j=1}^{N}\gamma_{j}\Bigg[\left(1+\bar{n}_{j}^{m}\right)\mathcal{L}_{\hat{b}_j}[\hat{\rho}]+\bar{n}_{j}^{m}\,\mathcal{L}_{\hat{b}^\dagger_j}[\hat{\rho}]\Bigg].\label{eq_master_equation}
\end{eqnarray}
In the above expression, $\mathcal{L}_{\hat{\mathcal{O}}}[\hat{\rho}] = \hat{\mathcal{O}}\hat{\rho}\hat{\mathcal{O}}^{\dagger}-\frac{1}{2}\big(\hat{\mathcal{O}}^{\dagger}\hat{\mathcal{O}}\hat{\rho}+\hat{\rho}\hat{\mathcal{O}}^{\dagger}\hat{\mathcal{O}}\big)$ accounts for the Lindblad superoperator describing the dissipative channel associated with operator $\hat{O}$. The parameters $\Gamma_{j}$ and $\gamma_j$ are the spin decay and mechanical damping rates, respectively. For simplicity we consider $\Gamma_{j}{=}\Gamma$ and $\gamma_j{=}\gamma$. Lastly, $\bar{n}_{j}^{s}$ ($\bar{n}_{j}^{m}$) specifies the mean thermal occupation of the spin (mechanical) reservoir at site $j$. For our simulations, without loss of generality, we initialize the spin systems as $\lvert \psi_0 \rangle{=}|{\downarrow}\rangle^{\bigotimes N}$ and the MOs in thermal states with $\bar{n}_{j}^{m}{=}0.1$. 

\section{Phonon lasing from elemental spin-mechanical system }
\subsection{Resonance Conditions for Phonon Lasing} To derive the resonance conditions and clarify the physical mechanism of phonon lasing, we first examine a minimal setup: two spins with driven coupling, where only one spin couples to a MO. The simplified Hamiltonian reads:
\begin{eqnarray}
\hat{\mathcal{H}}=\omega_1\hat{b}_1^{\dagger}\hat{b}_1{+}J_{1} \cos{(\Omega_1t)}\hat{\sigma}_{1}^{x}\hat{\sigma}_{2}^{x}{-}\lambda_{1}\hat{\sigma}_{1}^{z}(\hat{b}_1^{\dagger}{+}\hat{b}_1){+}\sum_{j=1}^2 \frac{\Delta_j}{2}\hat{\sigma}_{j}^{z}.~\label{eq_simplified_hamiltonian}
\end{eqnarray}
Eq.~\eqref{eq_simplified_hamiltonian} represents the simplest possible scenario in which lasing can emerge. Interestingly, by applying the rotating-wave approximation and weak spin-mechanical coupling, one can derive an effective Hamiltonian given by---see Supplemental Material (SM)~\cite{Supplemental} for details:
\begin{equation} \label{eq:heff1}
    \hat{\mathcal{H}}^{\mathrm{eff}}=i\frac{J_1\lambda_1}{\omega_1}\Bigg(\hat{\sigma}_{1}^{+}\hat{\sigma}_{2}^{+}\hat{b}_1^{\dagger}-h.c.\Bigg).
\end{equation}
Notably, the above spin-spin-phonon blue-sideband processes, such as $\hat{\sigma}_{1}^{+}\hat{\sigma}_{2}^{+}\hat{b}_1^{\dagger}$, have been selectively enabled by tuning the system to specific resonance conditions, see SM~\cite{Supplemental} for details:
\begin{equation}\label{eq_resonance_conditions}
\Omega_1 = \Delta_1+ \Delta_2+\omega_1.
\end{equation}
Note that other spin-phonon processes can also be enabled with another resonant condition resulting in phase-locked lasing. Although physically interesting, the underlying mechanism remains equivalent to the lasing with random phase, see SM~\cite{Supplemental} for details. Without loss of generality, we restrict our analysis to the case $\Delta_1=\Delta_2=2\overline \omega$. Moreover, all frequencies are normalized to $\overline\omega$, and time is expressed in dimensionless units as $\overline\omega t$.
\begin{figure}[t]
\centering
\includegraphics[width=1\linewidth]{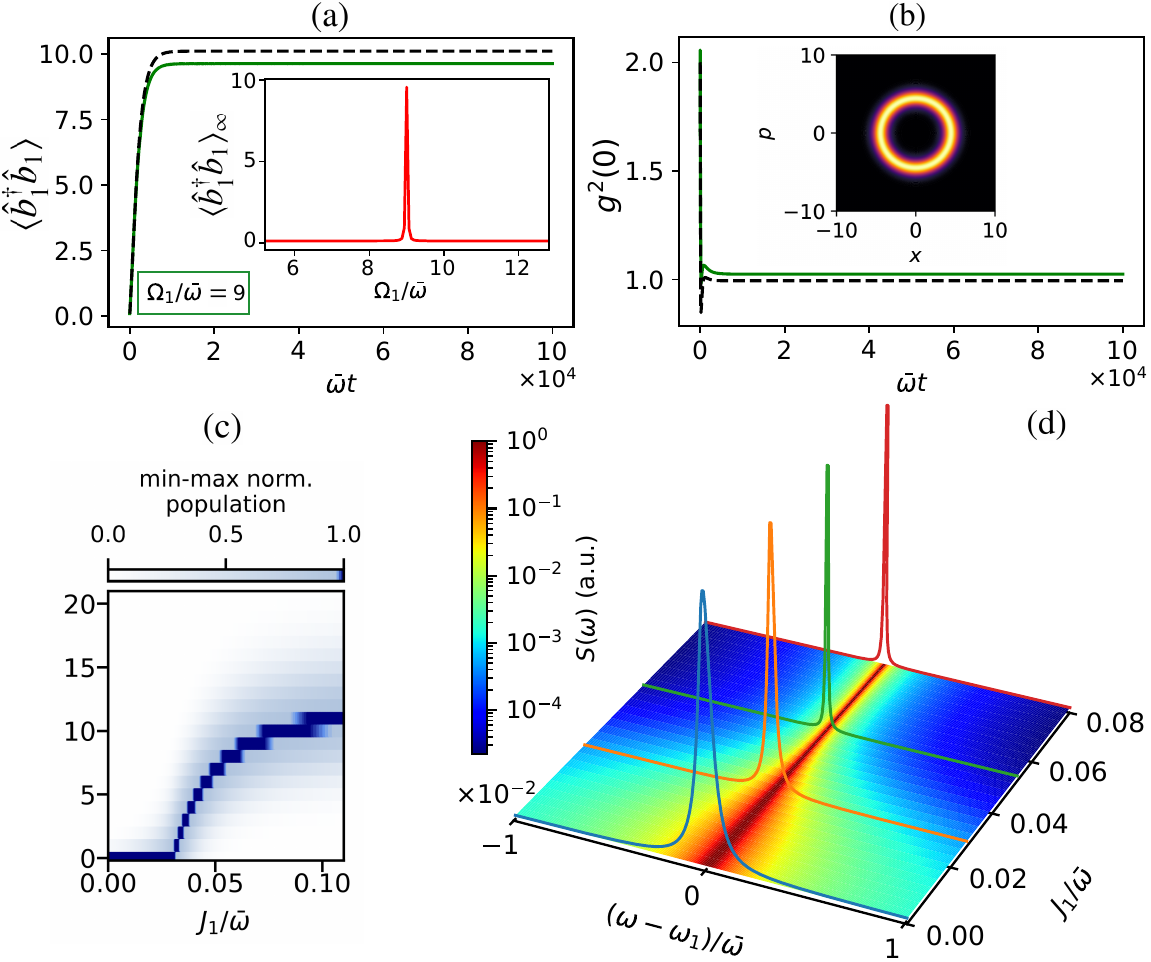}
\caption{Case of two spins, with only one spin coupled to a MO at resonance condition $\Omega_1=4\overline\omega+\omega_1$ ~\ref{eq_resonance_conditions}. (a) Comparison of the time evolution of phonon amplification, characterized by the average phonon number $\langle\hat{b}_1^\dagger \hat{b}_1\rangle$, for the full Hamiltonian (solid green line) and the effective Hamiltonian (dashed black line). Inset: shows the steady-state average phonon number $\langle\hat{b}_1^\dagger \hat{b}_1\rangle_\infty$ as a function of $\Omega_1/\overline\omega$. (b) Second-order correlation function $g^{(2)}(0)$ versus time $\overline\omega t$. Inset: steady-state $g^{(2)}(0)$ as a function of $\Omega_1/\overline\omega$, and Wigner distribution $W(x,p)$. We compare full Hamiltonian dynamics (solid green) with effective Hamiltonian results (dashed black). (c) Min-max normalized steady-state phonon population versus $J_1/\overline\omega$. (d) Power spectrum $S(\omega)$ as a function of $J_1/\overline\omega$. Frequency defined with respect to the MO frequency $\omega_1$. Other parameters are: $\Delta_1=\Delta_2=2\overline\omega$, $\omega_1=5\overline\omega$, $\Omega=9\overline\omega$, $\lambda_1=0.4\overline\omega$, $J_1=0.08\overline\omega$, $\Gamma_{1}=\Gamma_{2}=2\times10^{-2}\overline\omega$, $\gamma_1=8\times10^{-4}\overline\omega$, $\bar{n}_{1}^{s}=\bar{n}_{2}^{s}=0.01$ and $\bar{n}_{1}^{m}=0.1$.}
\label{fig_elemental_lasing}
\end{figure}

\subsection{Signatures of Phonon Lasing} 
To explicitly examine whether the amplification of phonon excitations saturates over time when losses are included, we numerically solve the master equation of Eq.~\eqref{eq_master_equation} for both the full and effective Hamiltonians under driven coupling conditions satisfying $J_1 \gg (\Gamma_1, \gamma_1)$, obtaining solutions $\rho(t)$ and $\rho^{\mathrm{eff}}(t)$, respectively. In Fig.~\ref{fig_elemental_lasing}\hyperref[fig_elemental_lasing]{(b)}, we plot the average phonon number $\langle \hat{b}_1^\dagger \hat{b}_1\rangle = \mathrm{Tr}[\rho(t)\hat{b}_1^\dagger \hat{b}_1]$ as a function of $\overline\omega t$ at the resonant frequency $\Omega_1 = 4\overline\omega+\omega_1$ (solid green line). For comparison, we also plot the effective model result $\langle \hat{b}_1^\dagger \hat{b}_1\rangle = \mathrm{Tr}[\rho^{\mathrm{eff}}(t)\hat{b}_1^\dagger \hat{b}_1]$ (dashed black line). Both numerical solutions closely match, clearly demonstrating saturation of phonon amplification---the second critical ingredient for phonon lasing. Indeed, the combined phonon amplification and saturation, resulting from the balance between gain and loss, indicate the emergence of phonon lasing. Furthermore, using the solution $\rho_{\mathrm{ss}} = \rho(\overline\omega t \gg 1)$ obtained numerically from the full Hamiltonian master equation, we plot the steady-state phonon number $\langle \hat{b}_1^\dagger \hat{b}_1\rangle_{\infty}$ vs. $\Omega_1/\overline\omega$ in the inset of Fig.~\ref{fig_elemental_lasing}\hyperref[fig_elemental_lasing]{(a)}. As clearly evidenced, sharp phonon amplification occurs precisely at the resonance condition $\Omega_1=4\overline\omega+\omega_1=9\overline\omega$, for $\omega_1=5\overline\omega$, consistent with predictions from our effective Hamiltonian analysis.

 As yet another signature of the generated phonon laser, we evaluate the second-order correlation function $g^{(2)}(0)=\frac{\langle\hat{b}^{\dagger 2} \hat{b}^{2}\rangle}{\langle\hat{b}^{\dagger}\hat{b}\rangle^{2}}$~\cite{glauber1963quantum}. In Fig.~\ref{fig_elemental_lasing}\hyperref[fig_elemental_lasing]{(b)}, we plot $g^{(2)}(0)$ as a function of time $\overline\omega t$ at the resonance frequency $\Omega_1=4\overline\omega+\omega_1=9\overline\omega$, comparing results from the full Hamiltonian dynamics (solid green curve) with those from the effective Hamiltonian (dashed black curve). As clearly seen from the figure, the second-order correlation function dynamically transitions from a thermal distribution $g^{(2)}(0)=2$ to a coherent state $g^{(2)}(0)\sim 1$. In the inset of Fig.~\ref{fig_elemental_lasing}\hyperref[fig_elemental_lasing]{(b)}, we present the Wigner quasiprobability distribution $W(x, p)=\frac{1}{\pi\hbar}\int_{-\infty}^{\infty}\langle x+y| \rho_{\text{oscillator}}|x - y\rangle e^{-2ipy/\hbar}dy$, for the oscillator subsystem's steady state, plotted in phase-space coordinates $(x, p)$ at the resonance condition $\Omega_1=9\overline\omega$. The distribution clearly illustrates the dynamical generation of a coherent state with random phase.

\subsection{Characterization of Phonon Lasing} 
To further characterize the stimulated phonon emission, we study the threshold condition for lasing emergence. In Fig.~\ref{fig_elemental_lasing}\hyperref[fig_elemental_lasing]{(c)}, we plot the min-max normalized phonon population obtained from the full dynamics of the oscillator's steady state as a function of the driving amplitude $J_1/\overline\omega$ for $\Omega_1= 4\overline\omega +\omega_1=9\overline\omega$. The figure clearly shows a distinct threshold around $J_1/\overline\omega \sim 0.03$, where the phonon population abruptly transitions from zero to non-zero values---that is, from spontaneous to stimulated phonon emission.

Finally, we numerically compute the phonon power spectrum in the frequency domain~\cite{gardiner2004quantum}, defined as $S(\omega)\equiv\int_{-\infty}^{+\infty}d\tau e^{-i\omega\tau} \langle\hat{b}^{\dagger}(\tau)\hat{b}(0)\rangle_\infty$. In Fig.~\ref{fig_elemental_lasing}\hyperref[fig_elemental_lasing]{(d)}, we present the power spectrum $S(\omega)$ as a function of $J_1/\overline{\omega}$ and $(\omega-\omega_1)/\overline{\omega}$ for the resonance condition $\Omega_1=9\overline\omega$. The spectra exhibit two clearly distinct regimes separated by the lasing threshold. Below threshold, i.e., for $J_1/\overline{\omega}\lesssim0.03$, the linewidth remains broad due to dominant thermal phonon fluctuations. Above threshold, a pronounced linewidth narrowing emerges, signaling the onset of coherent phonon lasing.

\section{Scalable Phonon Laser Arrays} 
\subsection{On-Demand Creation of Many-Body Phonon Lasers} 
Having characterized phonon lasing in a minimal setup, we now demonstrate scalable, on-demand arrays of $N$ phonon lasers. To precisely identify the resonance conditions that enable coherent amplification of MOs, we analytically derive an effective Hamiltonian for the general scenario of $N$ oscillators described by Eq.~\eqref{eq_driven_hamiltonian}. See SM~\cite{Supplemental}, e.g. the phase-locked coherent state, Eq. (S10):
\begin{equation}\label{eq:heff2}
    \hat{\mathcal{H}}_N^{\textit{eff}}=\sum_{j=1}^{N-1}\frac{J_j}{2}\hat{\sigma}_j^{+}\hat{\sigma}_{j+1}^{+}-i\sum_{j=1}^{N-1}\sum_{k=j}^{j+1}\frac{J_j \lambda_k}{\omega_k}\hat{\sigma}_j^{+}\hat{\sigma}_{j+1}^{+}\hat{b}_k^{\dagger} + h.c.
\end{equation}
Eq.~\eqref{eq:heff2} is the central expression of our work, as it precisely defines the blue-sideband resonance conditions for the most general case. Specifically, these resonance conditions become:
\begin{equation}
    \Omega_{j} = \sum_{k=j}^{j+1}\Delta_k \hspace{1cm} \text{and} \hspace{1cm} \omega_{k} =  2\Omega_{j}.
\end{equation}
To investigate the coupled dynamics of the spins and MOs, we solve the full set of Heisenberg equations derived from the time-dependent Hamiltonian in Eq.~\eqref{eq_driven_hamiltonian}. Although the complete set of equations can be analytically derived, see the SM~\cite{Supplemental} for details, here we present a minimal subset to clearly illustrate the interplay between phonon number and spin populations:
\begin{eqnarray}
\frac{d\langle \hat{b}_j^\dagger \hat{b}_j \rangle}{dt}&{=}& i\lambda_j\langle \hat{\sigma}_j^z \rangle(\langle \hat{b}_j^\dagger \rangle - \langle \hat{b}_j \rangle) - \gamma\langle \hat{b}_j^\dagger \hat{b}_j \rangle, \label{lang_f} \\
\frac{d\langle \hat{\sigma}_j^z \rangle}{dt}&{=}&2i J_{j-1} \cos{(\Omega_{j-1}t)}( \langle \hat{\sigma}_{j-1}^+ \hat{\sigma}_j^- \rangle + \langle \hat{\sigma}_{j-1}^- \hat{\sigma}_j^- \rangle \\
\nonumber &-& \langle \hat{\sigma}_{j-1}^+ \hat{\sigma}_j^+ \rangle - \langle \hat{\sigma}_{j-1}^- \hat{\sigma}_j^+ \rangle) + 2i J_j \cos{(\Omega_j t)} ( \langle \hat{\sigma}_j^- \hat{\sigma}_{j+1}^+ \rangle \\
\nonumber &+& \langle \hat{\sigma}_j^- \hat{\sigma}_{j+1}^- \rangle - \langle \hat{\sigma}_j^+ \hat{\sigma}_{j+1}^+ \rangle
- \langle \hat{\sigma}_j^+ \hat{\sigma}_{j+1}^- \rangle) - \Gamma( \langle \hat{\sigma}_j^z \rangle + 1).
\label{lang_f_1}
\end{eqnarray}
From the above equations, it is clear that the phonon number occupancy $\langle \hat{b}_j^\dagger \hat{b}_j \rangle$ at site $j$ dynamically depends on the spin inversion $\langle \hat{\sigma}_j^z \rangle$ at the same site. Importantly, Eq.~\eqref{lang_f_1} shows explicitly that the dynamics of this spin inversion are governed by spin-spin correlations between adjacent sites. Thus, to achieve phonon amplification, precise tuning and coherent control of neighbouring spin states are essential.

\begin{figure*}[t]
\centering
\includegraphics[width=1\linewidth]{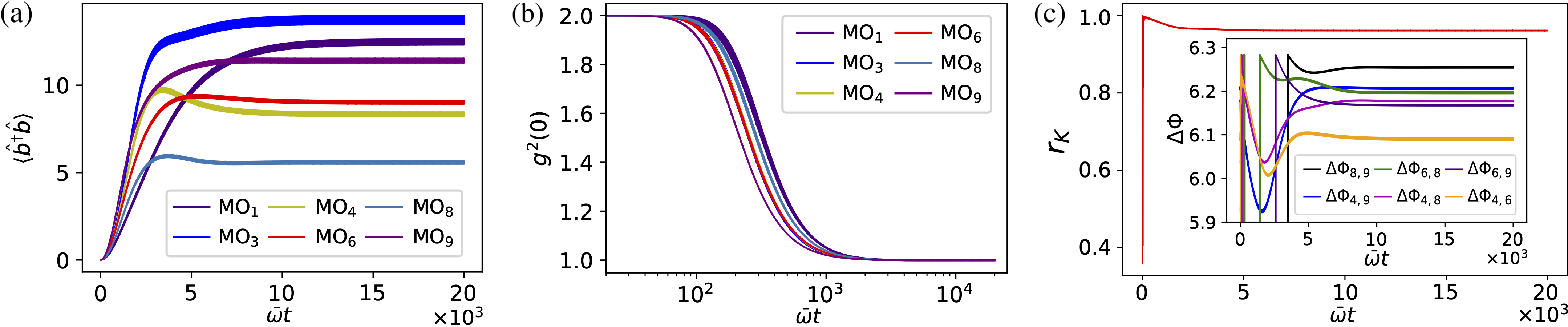}
\caption{Spin-mechanical array of $N=10$ sites, with MOs placed only at selected sites $j=(1,3,4,6,8,9)$. (a) Phonon expectation value $\langle \hat{b}_j^\dagger \hat{b}_j \rangle$ as a function of time $\overline\omega t$. (b) Second-order correlation function $g^{(2)}(0)$ of each MO as a function of time $\overline\omega t$ under the same resonance conditions as in (a). (c) Kuramoto parameter and (inset) phase difference for some pairs of lasing MOs. Other parameters are: $\omega_j/\overline\omega=\{8.0, 8.0, 7.9995, 7.9994, 8.0, 7.9992, 8.0, 7.999, 7.9989, 7.9988\}$, $\lambda_j/\overline\omega = \{0.4,0,0.42,0.38,0,0.41,0,0.37,0.43,0\}$, $J_j/\overline\omega=0.2$, $\Omega_j/\overline\omega=4$, $\Gamma/\overline\omega=8\times10^{-2}$, $\gamma/\overline\omega=10^{-3}$, $\bar{n}_{j}^{s}=0.01$ and $\bar{n}_{j}^{m}=0.1$.}
\label{fig_sync}
\end{figure*}

To illustrate the generality of our proposal, we explicitly consider a spin-mechanical array of $N=10$ sites, with MOs placed only at selected sites $j=(1,3,4,6,8,9)$. Any other configurations are equally feasible. As a first clear signature of phonon-laser arrays, Fig.~\ref{fig_sync}\hyperref[fig_sync]{(a)} shows the phonon expectation value $\langle \hat{b}_j^\dagger \hat{b}_j \rangle$ of each MO as a function of time $\overline\omega t$ for resonance conditions $\Omega_j/\overline\omega=4$ and $\omega_j/\overline\omega=8(1-\varepsilon)$ with $\varepsilon\ll 1$. As clearly shown in Fig.~\ref{fig_sync}\hyperref[fig_sync]{(a)}, each oscillator reaches its own saturation plateau, explicitly demonstrating individually addressable phonon amplification. In agreement with the mechanism described by the general effective Hamiltonian of Eq.~\eqref{eq:heff2}. Additional characterization is provided in Fig.~\ref{fig_sync}\hyperref[fig_sync]{(b)}, where we plot the second-order correlation function $g^{(2)}(0)$ of each MO as a function of time $\overline\omega t$ under the same resonance conditions. As clearly seen from the figure, all MOs exhibit $g^{(2)}(0)\sim 1$, confirming individually addressable coherent phonon amplification through stimulated emission.

\subsection{Self-Organized Synchronization of Phonon Lasers} 
To demonstrate the robustness of our phonon‑laser array, we examine signatures of self‑organized synchronization---a collective phenomenon widely studied in laser arrays~\cite{mari2013measures}. We first quantify global synchronization with the Kuramoto order parameter $r_K \equiv \frac{1}{M}\sum_{j\in N} e^{i\phi_j(t)}$, where $M\leq N$ is the number of active (lasing) MOs in the set of $N$ [in our current case $j=(1,3,4,6,8,9)$], and $\phi_j(t)$ is the instantaneous phase of oscillator $\phi_j=\arg\langle \hat b_j\rangle$ at site $j$. Note that $r_K{=}1$ indicates full global phase locking, whereas $r_K{=}0$ indicates no global synchronization~\cite{acebron2005kuramoto}. In Fig.~\ref{fig_sync}\hyperref[fig_sync]{(c)} we plot the Kuramoto parameter $r_K$ as a function of time under the same resonance conditions as those shown in Figs.~\ref{fig_sync}\hyperref[fig_sync]{(a)}-\hyperref[fig_sync]{(b)}. As clearly illustrated, full global synchronization occurs when MOs, initially prepared in distinct states, evolve dynamically into identical phase trajectories due to their mutual coupling, leading to $r_K \sim 1$. Additionally, we evaluate the pairwise instantaneous phase difference between oscillators $j$ and $k$, defined as $\Delta \Phi_{j,k}(t)=\arg(\langle\hat{b}_j(t)\rangle)-\arg(\langle\hat{b}_{k}(t)\rangle))$. In the inset of Fig.~\ref{fig_sync}\hyperref[fig_sync]{(c)}, we plot $\Delta \Phi_{j,k}(t)$ as a function of time $\overline\omega t$. After sufficient evolution time, the MOs become pairwise synchronized, and their oscillation amplitudes stabilize. As clearly shown in the inset of Fig.~\ref{fig_sync}\hyperref[fig_sync]{(c)}, pairwise phase synchronization occurs as $\Delta \Phi_{j,k}(t)$ asymptotically approaches a constant value within the interval $[0,2\pi]$~\cite{sheng2020selforganized, mari2013measures, hush2015spin}.

\subsection{Experimental Feasibility} One crucial ingredient of our scheme is the spin-mechanical interaction described by $\propto \hat{\sigma}_j^z(\hat{b}_j^\dagger+\hat{b}_j)$ in Eq.~\eqref{eq_driven_hamiltonian}. This interaction has already been experimentally realized on several platforms~\cite{kolkowitz2012coherent, arcizet2011single, lahaye2009nanomechanical, martinetz2020quantum, pirkkalainen2013hybrid, aporvari2021strong, pirkkalainen2015cavity}. Another crucial ingredient is the pair-raising (lowering) mechanism acting on the spins, which enables phonon lasing and their synchronization. This mechanism could be achieved in state-of-the-art quantum platforms by locally driving the spin gaps. In particular, a local driving of the form $H_0(t)=\hbar\varepsilon_0\cos(\nu t)\sigma^z_j$ acting upon the spin $j$ \cite{Dunlap1986, DunlapKenkre1988, Bukov2015AdvPhys} represents the simplest protocol to achieve spin pair-raising and can be experimentally implemented in superconducting circuits \cite{Zhao2022,Rosen2024,Song2024}. Locally driving spin gaps effectively renormalizes the spin-spin interaction strength by introducing first-order Bessel functions such that $J_j\to J_j\sum_{m\in \mathbb{Z}}\mathcal{J}_m(x)e^{\pm i m\nu t}$, see SM~\cite{Supplemental} for details. If the driving frequency $\nu=12\overline\omega$ ($\nu=2\overline\omega$), $\omega_1=8\overline\omega$, with $\overline\omega\gg J_1\lambda_1/\omega_1$, it leads to the Case I (II) described in the SM~\cite{Supplemental}, activating the desired spin–phonon blue-sideband processes, such as $\hat{\sigma}_{1}^{+}\hat{\sigma}_{2}^{+}\hat{b}_1^{\dagger}$. 

Interestingly, we believe that a possible physical implementation of our proposal can be realized within circuit quantum acoustic dynamics \cite{chu2017quantum}. Some experiments have already shown the coupling of an acoustic resonator and a superconducting transmon qubit \cite{potts2025large}. Nowadays, superconducting devices such as the fluxonium can be engineered in coplanar architectures and exhibit frequency gaps of about $\overline\omega=2\pi\times 1.09$ GHz \cite{bao2022fluxonium} and $\overline\omega=2\pi\times 0.2$ GHz \cite{ding2023high}. Furthermore, fluxonium devices can be integrated with a mechanical oscillator exhibiting a strong dispersive coupling \cite{lee2023strong}. Based on those low-frequency superconducting qubits, we estimate resonance frequencies [see Eq.~\eqref{eq_resonance_conditions}] for the pair-raising (lowering) phenomenon of about $\Omega_1=2\pi\times 4.36$ GHz and $\omega_1=2\pi\times 8.72$ GHz for the first fluxonium, whereas $\Omega_1=2\pi\times 0.8$ GHz and $\omega_1=2\pi\times 1.6$ GHz for the second fluxonium. These frequency values are well within the current technology. For instance, mechanical oscillators with resonance frequencies of $2\pi \times 6.8$ GHz and $2\pi \times 5.2$ GHz have been reported in Refs.~\cite{ng2023intermodulation,madiotoptomechanical2023,mirhosseini2020superconducting}. Additionally, driving frequencies acting upon superconducting circuits are available within the MHz-GHz range \cite{manucharyan2009fluxonium,macklin2015near}. Given the discussion above, we firmly believe that circuit quantum acoustic represents a strong candidate for implementing scalable phonon-laser arrays.

\section{Conclusions}
In this work, we propose a scheme for the scalable generation of phonon-laser arrays through local control. We demonstrate the flexibility of our approach by considering a quantum many-body spin chain with Ising-like interactions under local time-dependent driving, where each mechanical oscillator (MO) couples individually to a single spin. Notably, we analytically derive a general effective Hamiltonian by selecting spin-phonon blue-sideband transitions via specific resonance conditions. With this effective Hamiltonian, we show that our scheme is universally applicable and valid for any spin chain length and any number of MOs. To support our findings, we evaluate the phonon number amplification and the second-order correlation function, both confirming coherent mechanical amplification via stimulated emission. Furthermore, our phonon-laser array is robust against small deviations in resonance conditions, exhibiting self-organized synchronization characterized both by global synchronization and pairwise phase-locking. Finally, we envision that our proposal can be physically implemented within circuit quantum acoustic dynamics.

Since our scheme enables scalable and on-demand phonon lasing at arbitrary lattice sites, we believe our results pave the way for advancements in quantum technologies, long-distance synchronization protocols, and quantum-acoustic many-body phenomena.

\section*{Acknowledgments}  H.M. acknowledges Universidad de La Frontera and partial financial support from the project ``FRO2395”, from the Ministry of Education of Chile. G.R. acknowledges the Center for Nanoscience and Nanotechnology, CEDENNA, Project CIA250002, and Proyecto de Exploraci\'on Grant No. 13250014. V.M. thanks support from the National Natural Science Foundation of China Grants No. 12374482 and No. W2432005. V.E. acknowledges support from ANID Fondecyt Regular No. 1221250.


\bibliography{biblio}

\clearpage
\onecolumngrid

\setcounter{tocdepth}{2}
\setcounter{section}{0}

\renewcommand{\thesection}{S\arabic{section}}
\renewcommand{\thesubsection}{S\arabic{section}.\arabic{subsection}}

\setcounter{figure}{0}
\renewcommand{\thefigure}{S\arabic{figure}}
\renewcommand{\theequation}{S\arabic{equation}}
\setcounter{equation}{0}

\section*{Supplementary Material}\label{Supplemental material}


\section{General Effective Hamiltonian for Phonon Laser}
\label{AppendA}
We begin by deriving the effective Hamiltonian applicable to an arbitrary number $N$ of spins and mechanical oscillators (MO). We then analyse the fundamental configuration of two spins coupled to a single MO, which serves as the example studied in the main text. For a system of $N$ spins coupled to $N$ MOs the Hamiltonian is given by ($\hbar=1$): 
\begin{equation} \label{eq:Hsync}
\hat{\mathcal{H}} =\sum_{j=1}^N \big[\frac{\Delta_j}{2} \hat{\sigma}_{j}^{z} + \omega_j\hat{b}_j^{\dagger}\hat{b}_j - \lambda_{j}\hat{\sigma}_{j}^{z}(\hat{b}_j^{\dagger}+\hat{b}_j)\big]+ \sum_{j=1}^{N-1}J_{j} \cos{(\Omega_jt)} \hat{\sigma}_{j}^{x}\hat{\sigma}_{j+1}^{x}
\end{equation}

To derive the effective Hamiltonian we begin by evaluating the Hamiltonian in the first interaction picture: $\hat{\mathcal{V}}=e^{i \hat{\mathcal{H}}_{0} t}\hat{\mathcal{H}}_{1}e^{-i \hat{\mathcal{H}}_{0} t}$, where 
\begin{eqnarray}
\hat{\mathcal{H}}_{0}&=&\sum_{j=1}^N \frac{\Delta_j}{2} \hat{\sigma}_{j}^{z} + \omega_j\hat{b}_j^{\dagger}\hat{b}_j, \\ \hat{\mathcal{H}}_{1}&=&\sum_{j=1}^{N-1}J_{j} \cos{(\Omega_jt)} \hat{\sigma}_{j}^{x}\hat{\sigma}_{j+1}^{x}-\sum_{j=1}^N \lambda_{j}\hat{\sigma}_{j}^{z}(\hat{b}_j^{\dagger}+\hat{b}_j).
\end{eqnarray}

With the above Hamiltonians and interaction picture transformation, one gets: $\hat{\mathcal{V}}=\hat{\mathcal{V}}_{0}+\hat{\mathcal{V}}_{1}$, where
\begin{align}
    \hat{\mathcal{V}}_{0}&=-i \sum_{j=1}^N\lambda_j\,\hat{\sigma}_{j}^{z}\,\left(\hat{b}_j^{\dagger}\,e^{i\omega_j t}-\hat{b}_j\,e^{-i\omega_j t}\right)\equiv\sum_{j=1}^N\hat{\sigma}_{j}^{z}\,\hat{f}_j(t),\\
    \hat{\mathcal{V}}_{1}&=\sum_{j=1}^{N-1}J_j\,\cos(\Omega_j t)\Big(e^{i(\Delta_j+\Delta_{j+1})t} \hat{\sigma}_j^{+} \hat{\sigma}_{j+1}^{+} + e^{i(\Delta_j-\Delta_{j+1})t} \hat{\sigma}_j^{+} \hat{\sigma}_{j+1}^{-}\Big)+ H.c,
\end{align}
with $\hat{f}_j(t)=-i\lambda_j\,\left(\hat{b}_j^{\dagger}\,e^{i\omega_jt}-\hat{b}_j\,e^{-i\omega_jt}\right)$.

Next, we move to a second interaction picture, which is defined as follows:
\begin{eqnarray}\label{2nd}    \hat{\mathcal{V}}'&=&\exp{\left\{i\int\hat{\mathcal{V}}_{0}dt\right\}}\hat{\mathcal{V}}_{1}\exp{\left\{-i\int\hat{\mathcal{V}}_{0}dt\right\}}\nonumber\\
    &=&\sum_{j=1}^{N-1}J_j\,\cos(\Omega_j t)\Big[e^{i(\Delta_j+\Delta_{j+1})t}e^{i\sum_{j=1}^{N}\hat{\sigma}_{j}^{z}\,\hat{F}_j(t)}\hat{\sigma}_j^{+} \hat{\sigma}_{j+1}^{+}e^{-i\sum_{j=1}^{N}\hat{\sigma}_{j}^{z}\,\hat{F}_j(t)}+e^{i(\Delta_j-\Delta_{j+1})t}e^{i\sum_{j=1}^{N}\hat{\sigma}_{j}^{z}\,\hat{F}_j(t)}\hat{\sigma}_{j}^{+} \hat{\sigma}_{j+1}^{-}e^{-i\sum_{j=1}^{N}\hat{\sigma}_{j}^{z}\,\hat{F}_j(t)}\Big] + H.c. \nonumber\\
    &\equiv& \hat{\mathcal{V}}'_{I}+\hat{\mathcal{V}}'_{II}+ H.c.,
\end{eqnarray}
with the Hermitian operators $\hat{F}_j(t)\equiv\int \hat{f}_j(t)\, dt=\frac{\lambda_j}{\omega_{j}}\left(\hat{b}_j^{\dagger}\eta_j+\hat{b}_j\eta_j^{*}\right)$, here $\eta_j=e^{i\omega_{j}t}-1$. We note that although $\hat{\mathcal V}_0(t)$ is time dependent, the time-ordering operator is not required in the evolution operator. This follows from the structure $\hat{\mathcal V}_0(t)=\sum_j \hat{\sigma}_j^z \hat f_j(t)$, together with the bosonic commutation relations $[\hat b_j,\hat b_k^\dagger]=\delta_{jk}$. In particular, one finds that $[\hat f_j(t),\hat f_k(t')]=0$ for $j\neq k$, while $[\hat f_j(t),\hat f_j(t')]=2i\lambda_j^2\sin[\omega_j(t-t')]$ is a $c$-number. Consequently, the commutator $[\hat{\mathcal V}_0(t),\hat{\mathcal V}_0(t')]$ is proportional to the identity operator, and the time-ordering operator contributes only a global phase. Therefore the evolution operator can be written as $\exp(i\int \hat{\mathcal V}_0 dt)$ without affecting the dynamics.

In deriving the effective Hamiltonian, the central approximation is that the spin–mechanical coupling strength $\lambda_j$ is considerably smaller than the mechanical frequency $\omega_j$, thereby permitting the following analytical simplification: $e^{i \hat{F}_j(t)}\approx1+i\frac{\lambda_j}{\omega_{j}}\left(\hat{b}_{j}^{\dagger}\eta_{j}+\hat{b}_{j}\eta_{j}^{*}\right)$. 

In what follows, the explicit form of each operator $\hat{\mathcal{V}}'$ appearing in Eq.~\eqref{2nd} is derived:
\begin{eqnarray}
    \hat{\mathcal{V}}'_{I}&\equiv&\sum_{j=1}^{N-1}J_j\,\cos(\Omega_j te^{i(\Delta_j+\Delta_{j+1})t}e^{i\sum_{j=1}^{N}\hat{\sigma}_{j}^{z}\,\hat{F}_j(t)}\hat{\sigma}_j^{+} \hat{\sigma}_{j+1}^{+}e^{-i\sum_{j=1}^{N}\hat{\sigma}_{j}^{z}\,\hat{F}_j(t)}=\sum_{j=1}^{N-1}J_j\,\cos(\Omega_j t)e^{i(\Delta_j+\Delta_{j+1})t}\hat{\sigma}_j^{+} \hat{\sigma}_{j+1}^{+}e^{2i\hat{F}_j(t)}e^{2i\hat{F}_{j+1}(t)}\nonumber\\
    &=&\sum_{j=1}^{N-1}J_j\,\cos(\Omega_j t)e^{i(\Delta_j+\Delta_{j+1})t}\hat{\sigma}_j^{+} \hat{\sigma}_{j+1}^{+}\Big[1+2i\frac{\lambda_j}{\omega_{j}}\left(\hat{b}_j^{\dagger}\eta_j+\hat{b}_1\eta_j^{*}\right)\Big]\Big[1+2i\frac{\lambda_{j+1}}{\omega_{j+1}}\left(\hat{b}_{j+1}^{\dagger}\eta_{j+1}+\hat{b}_1\eta_{j+1}^{*}\right)\Big]\nonumber\\
    &\simeq&\sum_{j=1}^{N-1}J_j\,\cos(\Omega_j t)e^{i(\Delta_j+\Delta_{j+1})t}\hat{\sigma}_j^{+} \hat{\sigma}_{j+1}^{+}+2i\sum_{j=1}^{N-1}\sum_{k=j}^{j+1}\frac{J_j \lambda_k}{\omega_k}\,\cos(\Omega_j t)e^{i(\Delta_j+\Delta_{j+1})t}\hat{\sigma}_j^{+} \hat{\sigma}_{j+1}^{+}\left(\hat{b}_{k}^{\dagger}\eta_{k}+\hat{b}_k\eta_{k}^{*}\right)\nonumber\\
    &=&\sum_{j=1}^{N-1}\frac{J_j}{2}\,\left(e^{i(\Omega_j +\Delta_j+\Delta_{j+1})t}+e^{i(-\Omega_j+\Delta_j+\Delta_{j+1})t}\right)\hat{\sigma}_j^{+} \hat{\sigma}_{j+1}^{+}+2i\sum_{j=1}^{N-1}\sum_{k=j}^{j+1}\frac{J_j \lambda_k}{\omega_k}\Big[\hat{\sigma}_j^{+}\hat{\sigma}_{j+1}^{+}\hat{b}_k^{\dagger}\left(e^{i(\Omega_j+\Delta_j+\Delta_{j+1}+\omega_{k})t}+e^{i(-\Omega_j+\Delta_j+\Delta_{j+1}+\omega_{k})t}\right)\nonumber\\
    &-&\hat{\sigma}_j^{+}\hat{\sigma}_{j+1}^{+}\left(\hat{b}_k^{\dagger}+\hat{b}_k\right)\left(e^{i(\Omega_j+\Delta_j+\Delta_{j+1})t}+e^{i(-\Omega_j+\Delta_j+\Delta_{j+1})t}\right)+ \hat{\sigma}_j^{+}\hat{\sigma}_{j+1}^{+}\hat{b}_k\left(e^{i(\Omega_j+\Delta_j+\Delta_{j+1}-\omega_{k})t}+e^{i(-\Omega_j+\Delta_j+\Delta_{j+1}-\omega_{k})t}\right)\Big].
\end{eqnarray}
\begin{eqnarray}
    \hat{\mathcal{V}}'_{II}&\equiv&\sum_{j=1}^{N-1}J_j\,\cos(\Omega_j t)e^{i(\Delta_j-\Delta_{j+1})t}e^{i\sum_{j=1}^{N}\hat{\sigma}_{j}^{z}\,\hat{F}_j(t)}\hat{\sigma}_j^{+} \hat{\sigma}_{j+1}^{-}e^{-i\sum_{j=1}^{N}\hat{\sigma}_{j}^{z}\,\hat{F}_j(t)}=\sum_{j=1}^{N-1}J_j\,\cos(\Omega_j t)e^{i(\Delta_j-\Delta_{j+1})t}\hat{\sigma}_j^{+} \hat{\sigma}_{j+1}^{-}e^{2i\hat{F}_j(t)}e^{-2i\hat{F}_{j+1}(t)}\nonumber\\
    &=&\sum_{j=1}^{N-1}J_j\,\cos(\Omega_j t)e^{i(\Delta_j-\Delta_{j+1})t}\hat{\sigma}_j^{+} \hat{\sigma}_{j+1}^{-}\Big[1+2i\frac{\lambda_j}{\omega_{j}}\left(\hat{b}_j^{\dagger}\eta_j+\hat{b}_1\eta_j^{*}\right)\Big]\Big[1-2i\frac{\lambda_{j+1}}{\omega_{j+1}}\left(\hat{b}_{j+1}^{\dagger}\eta_{j+1}+\hat{b}_1\eta_{j+1}^{*}\right)\Big]\nonumber\\
    &\simeq&\sum_{j=1}^{N-1}J_j\,\cos(\Omega_j t)e^{i(\Delta_j-\Delta_{j+1})t}\hat{\sigma}_j^{+} \hat{\sigma}_{j+1}^{-}+2i\sum_{j=1}^{N-1}J_j \cos(\Omega_j t)e^{i(\Delta_j-\Delta_{j+1})t}\hat{\sigma}_j^{+} \hat{\sigma}_{j+1}^{-}\Big[\frac{\lambda_j}{\omega_j}\,\left(\hat{b}_{j}^{\dagger}\eta_{j}+\hat{b}_j\eta_{j}^{*}\right)-\frac{ \lambda_{j+1}}{\omega_{j+1}}\,\left(\hat{b}_{j+1}^{\dagger}\eta_{j+1}+\hat{b}_{j+1}\eta_{j+1}^{*}\right)\Big]\nonumber\\
    &=&\sum_{j=1}^{N-1}\frac{J_j}{2}\,\left(e^{i(\Omega_j+\Delta_j-\Delta_{j+1})t}+e^{i(-\Omega_j+\Delta_j-\Delta_{j+1})t}\right)\hat{\sigma}_j^{+} \hat{\sigma}_{j+1}^{-}+2i\sum_{j=1}^{N-1}J_j\Big[\frac{\lambda_j}{\omega_j}\hat{\sigma}_j^{+}\hat{\sigma}_{j+1}^{-}\hat{b}_j^{\dagger}\left(e^{i(\Omega_j+\Delta_j-\Delta_{j+1}+\omega_{j})t}+e^{i(-\Omega_j+\Delta_j-\Delta_{j+1}+\omega_{j})t}\right)\nonumber\\
    &-&\frac{\lambda_{j+1}}{\omega_{j+1}}\hat{\sigma}_j^{+}\hat{\sigma}_{j+1}^{-}\hat{b}_{j+1}^{\dagger}\left(e^{i(\Omega_j+\Delta_j-\Delta_{j+1}+\omega_{j+1})t}+e^{i(-\Omega_j+\Delta_j-\Delta_{j+1}+\omega_{j+1})t}\right)\nonumber\\
    &-&\frac{\lambda_j}{\omega_j}\hat{\sigma}_j^{+}\hat{\sigma}_{j+1}^{-}\left(\hat{b}_j^{\dagger}+\hat{b}_j\right)\left(e^{i(\Omega_j+\Delta_j-\Delta_{j+1})t}+e^{i(-\Omega_j+\Delta_j-\Delta_{j+1})t}\right)+\frac{\lambda_{j+1}}{\omega_{j+1}}\hat{\sigma}_j^{+}\hat{\sigma}_{j+1}^{-}\left(\hat{b}_{j+1}^{\dagger}+\hat{b}_{j+1}\right)\left(e^{i(\Omega_j+\Delta_j-\Delta_{j+1})t}+e^{i(-\Omega_j+\Delta_j-\Delta_{j+1})t}\right)\nonumber\\
    &+&\frac{\lambda_j}{\omega_j} \hat{\sigma}_j^{+}\hat{\sigma}_{j+1}^{-}\hat{b}_j\left(e^{i(\Omega_j+\Delta_j-\Delta_{j+1}-\omega_{j})t}+e^{i(-\Omega_j+\Delta_j-\Delta_{j+1}-\omega_{j})t}\right)-\frac{\lambda_{j+1}}{\omega_{j+1}} \hat{\sigma}_j^{+}\hat{\sigma}_{j+1}^{-}\hat{b}_{j+1}\left(e^{i(\Omega_j+\Delta_j-\Delta_{j+1}-\omega_{j+1})t}+e^{i(-\Omega_j+\Delta_j-\Delta_{j+1}-\omega_{j+1})t}\right)\Big].
\end{eqnarray}
Finally, Eq.~\eqref{2nd} can be significantly simplified within the framework of the rotating-wave approximation (RWA). In this approximation, only the non-oscillating (i.e. time-independent) contributions are retained. The deliberate enforcement of well‑defined resonance conditions enables the selective realization of two distinct lasing regimes. These regimes are characterized by (i) a coherent $\alpha$‑state emission profile and (ii) a doughnut‑shaped mode exhibiting an indeterminate phase distribution.

\begin{figure}[t]
\centering
\includegraphics[width=0.7\linewidth]{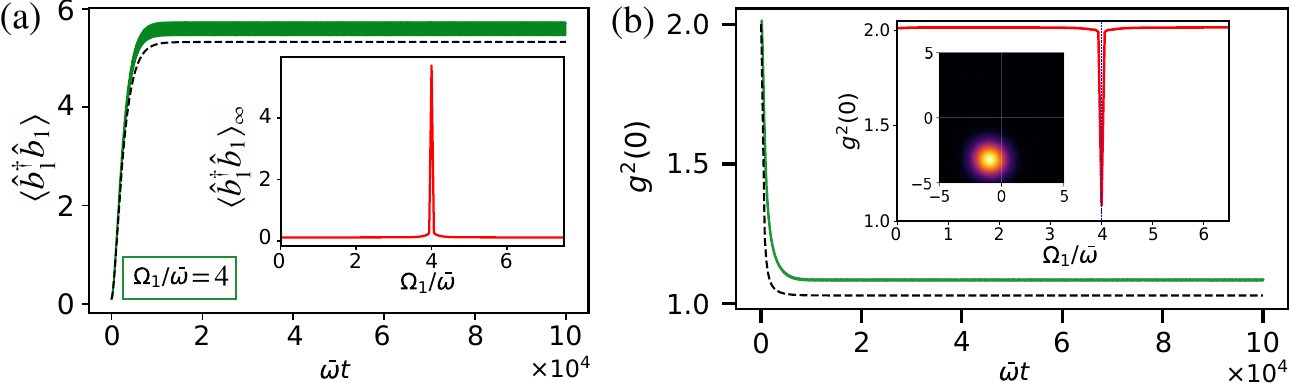}
\caption{(a) Time evolution of phonon amplification quantified by the average phonon number $\langle \hat{b}_1^\dagger \hat{b}_1 \rangle$ at the resonance condition $\Omega_1 = 4\overline\omega$. The inset shows the corresponding steady-state value $\langle \hat{b}_1^\dagger \hat{b}_1 \rangle_{\infty}$ as a function of $\Omega_1/\overline\omega$. (b) Second-order correlation function $g^{(2)}(0)$ versus dimensionless time $\overline\omega t$ at the same resonance. Inset: Wigner distribution $W(x,p)$ evaluated at $\Omega_1/\overline\omega = 8$. Solid green lines denote the full Hamiltonian dynamics, while dashed black lines correspond to the effective Hamiltonian, $\hat{\mathcal{H}}_{II}^{\textit{eff}}$. Other parameters are: $\Delta_1=\Delta_2=2\overline\omega$, $\omega_1=8\overline\omega$, $\lambda_1=0.4\overline\omega$, $J_1=0.1\overline\omega$, $\Gamma_{1}=\Gamma_{2}=8\times10^{-3}\overline\omega$, $\gamma_1=10^{-3}\overline\omega$, $\bar{n}_{1}^{s}=\bar{n}_{2}^{s}=0.01$ and $\bar{n}_{1}^{m}=0.1$.}
\label{fig_phase_locked}
\end{figure}

\textit{Case I.} Under the resonance conditions $\sum_{k=j}^{j+1}\Delta_k + \omega_{k} = \pm \Omega_{j}$, with $k=(j,j+1)$, with $\Delta_j\gg\left\{J_j,\lambda_j\right\}$, one gets the effective Hamiltonian for the phonon coherent state with a random phase, i.e. doughnut‑shaped lasing mode in Wigner representation:
\begin{equation}\label{eff2}
\hat{\mathcal{H}}_{I}^{\textit{eff}}=i\sum_{j=1}^{N-1}\sum_{k=j}^{j+1}\frac{J_j \lambda_k}{\omega_k}\hat{\sigma}_j^{+}\hat{\sigma}_{j+1}^{+}\hat{b}_k^{\dagger} + H.c.
\end{equation}

\textit{Case II.} Under the resonance conditions $\Omega_{j} = \sum_{k=j}^{j+1}\Delta_k$, $\omega_{k} = 2\Omega_{j}$ and $\Delta_k\gg\left\{J_j,\lambda_j\right\}$, with $k=(j,j+1)$, the system exhibits a phase-locked regime (see Fig. \ref{fig_phase_locked} for elemental model):
\begin{equation}\label{eff1}
    \hat{\mathcal{H}}_{II}^{\textit{eff}}=\sum_{j=1}^{N-1}\frac{J_j}{2}\hat{\sigma}_j^{+}\hat{\sigma}_{j+1}^{+}-i\sum_{j=1}^{N-1}\sum_{k=j}^{j+1}\frac{J_j \lambda_k}{\omega_k}\hat{\sigma}_j^{+}\hat{\sigma}_{j+1}^{+}\hat{b}_k^{\dagger} + H.c.
\end{equation}

\subsection*{Elemental model of two spins and one MO (analysed in the main text)}
\label{AppendA2}
This particular case arises as a reduction of the more general configuration involving \(N\) spins/MOs, focusing instead on the simpler setup of two spins coupled to a single MO, as presented in the main text. By concentrating on this minimal system, we isolate the core interaction mechanisms while preserving all essential physical features of the full model. For the system of two spins and one MO the Hamiltonian is given by ($\hbar=1$): 
\begin{equation} \label{eq:Helem}
\mathcal{\hat{H}}_{1}= \frac{\Delta_{1}}{2}\hat{\sigma}_1^z +\frac{\Delta_{1}}{2}\hat{\sigma}_2^z 
+  \omega_1\hat{b}_1^\dagger \hat{b}_1 
+ J_1\,\cos{(\Omega_1 t)}\,\hat{\sigma}_{1}^{x}\hat{\sigma}_{2}^{x}-\lambda\,\hat{\sigma}_{1}^{z}\,(\hat{b}_1^{\dagger}+\hat{b}_1)
\end{equation}

Consequently, the effective Hamiltonian for this elementary configuration takes the form:

\textit{(I) Random-phase lasing:} Under the resonance conditions $\sum_{j=1}^{2}\Delta_{j} + \omega_{1} = \pm \Omega_{1}$ one gets the effective Hamiltonian for the doughnut‑shaped mode laser:
\begin{equation}\label{effdona}
\hat{\mathcal{H}}_{I}^{\textit{eff}}=i\frac{J_1\lambda_1}{\omega_1}\hat{\sigma}_1^{+}\hat{\sigma}_2^{+}\hat{b}_1^{\dagger} + H.c.
\end{equation}

\textit{(II) Phase-locked regime:} Under the resonance conditions $\Omega_{1} = \sum_{j=1}^{2}\Delta_{j}$ and $\omega_{1} = 2\Omega_{1}$, one gets the effective Hamiltonian (discussed in the main text, Eq.4) for the phonon coherent state:
\begin{equation}\label{effcoh}
    \hat{\mathcal{H}}_{II}^{\textit{eff}}=\frac{J_1}{2}\hat{\sigma}_1^{+}\hat{\sigma}_2^{+} -i\frac{J_1\lambda_1}{\omega_1}\hat{\sigma}_1^{+}\hat{\sigma}_2^{+}\hat{b}_1^{\dagger} + H.c.
\end{equation}

As the effective Hamiltonian \eqref{effcoh} is already analysed in the main text, here we present the corresponding lasing characteristics for case (I), see Fig.~\ref{fig_phase_locked}.


\section{\label{AppendB} Heisenberg-Langevin equations for operator expectation values}

\subsection{General approach}
This section provides an expanded derivation of the kinetic equations governing the dynamics of the expectation values of the spin operators $\{\hat{\sigma}_j^+, \hat{\sigma}_j^-, \hat{\sigma}_j^z\}$ and the bosonic operators $\{\hat{b}^\dagger_j, \hat{b}_j\}$ for each mechanical oscillator $j = 1,...,N$. These equations follow directly from the quantum master equation (ME) associated with a general system Hamiltonian, and they form the basis for analysing the coupled spin–phonon dynamics. The ME for the joint spin-mechanical density operator $\hat{\rho}$ reads:
\begin{equation*}\label{eq:master-equation}
     \frac{d\hat{\rho}}{dt}=-i[\mathcal{\hat{H}},\hat{\rho}]+\sum_{j=1}^N \Gamma_{j} \mathcal{L}_{\hat{\sigma}_{j}^{-}}[\hat{\rho}] + \gamma_j \mathcal{L}_{\hat{b}_j}[\hat{\rho}],
\end{equation*}
with $\hat{\mathcal{H}}$ defined in Eq. \eqref{eq:Hsync}, 
$\mathcal{L}_{\hat{\mathcal{O}}}[\hat{\rho}] = \hat{\mathcal{O}}\hat{\rho} \,\hat{\mathcal{O}}^{\dagger}-\frac{1}{2}(\hat{\mathcal{O}}^{\dagger}\hat{\mathcal{O}}\hat{\rho}+\hat{\rho}\,\hat{\mathcal{O}}^{\dagger}\hat{\mathcal{O}})$ and $\Gamma_j$ ($\gamma_k$) representing the damping rate characterizing the interaction of each spin (MO) with its corresponding thermal reservoir, accounting for energy dissipation into the bath. Here we adopt the zero-temperature approximation $\bar{n}_{j}^{s} = \bar{n}_j^m=0$, valid for the spin and mechanical frequencies considered in our calculations (on the order $\sim$ GHz). At cryogenic temperatures around $10$ mK, thermal occupations are negligible. 

Therefore, the time evolution of the expectation value of any operator (spin or bosonic) $\hat{O}$ can be written as: $\frac{d\langle \hat{O} \rangle}{dt} = \text{Tr}\left(\hat{O} \frac{d\hat{\rho}}{dt}\right)$.

Substituting the master equation and using cyclic invariance of the trace, one obtains:

i) \textbf{Unitary part}: $\text{Tr}\left(\hat{O} (-i[\hat{\mathcal{H}}, \hat{\rho}])\right) = -i \langle [\hat{O}, \hat{\mathcal{H}}] \rangle$.

ii) \textbf{Dissipative part} (mechanical damping): $
   \sum_{j=1}^N \gamma_j\text{Tr}\left(\hat{O} \mathcal{L}_{\hat{b}_j}[\hat{\rho}]\right) = \sum_{j=1}^N\frac{\gamma_j}{2} \left( 2\langle \hat{b}_j^\dagger \hat{O} \hat{b}_j \rangle - \langle \hat{b}_j^\dagger \hat{b}_j \hat{O} \rangle - \langle \hat{O} \hat{b}_j^\dagger \hat{b}_j \rangle \right)$.

iii) \textbf{Dissipative part} (spin decay):  $
   \sum_{j=1}^N \Gamma_j\text{Tr}\left(\hat{O} \mathcal{L}_{\hat{\sigma}_j^-}[\hat{\rho}]\right) = \sum_{j=1}^N \frac{\Gamma_j}{2} \left( 2\langle \hat{\sigma}_j^+ \hat{O} \hat{\sigma}_j^- \rangle - \langle \hat{\sigma}_j^+ \hat{\sigma}_j^- \hat{O} \rangle - \langle \hat{O} \hat{\sigma}_j^+ \hat{\sigma}_j^- \rangle \right)$.
   
Combining all terms, we get the Heisenberg-Langevin equation for the mean value of any operator $\hat{O}$, which reads:
\begin{equation} \label{HLeq}
\frac{d\langle \hat{O} \rangle}{dt} = i \langle [\hat{\mathcal{H}}, \hat{O}] \rangle +  \sum_{j=1}^N \frac{\gamma_j}{2}\left( 2\langle \hat{b}_j^\dagger \hat{O} \hat{b}_j \rangle - \langle \hat{b}_j^\dagger \hat{b}_j \hat{O} \rangle - \langle \hat{O} \hat{b}_j^\dagger \hat{b}_j \rangle \right) + \frac{\Gamma_j}{2} \left( 2\langle \hat{\sigma}_j^+ \hat{O} \hat{\sigma}_j^- \rangle - \langle \hat{\sigma}_j^+ \hat{\sigma}_j^- \hat{O} \rangle - \langle \hat{O} \hat{\sigma}_j^+ \hat{\sigma}_j^- \rangle \right).
\end{equation}

\subsection{Closed set of equations for the array of $N$ spins and $N$ MOs} 

In the following, we derive the kinetic equations for the mean values of the spin and MO subsystem operators, together with the spin–spin correlations. The derivation is based on the time-dependent Hamiltonian \eqref{eq:Hsync} and makes use of Eq.~\eqref{HLeq}, under the simplifying assumptions  $\gamma_j\equiv\gamma $ and $\Gamma_j\equiv\Gamma $. Furthermore, provided the coupling constants satisfy $J_j\lambda_j/\omega_j \ll \{\Delta_j, J, \lambda\}$, we apply a mean-field decoupling: $\langle \hat{\sigma}_j^z \hat{\sigma}_{j\pm1}^\pm \rangle \approx \langle \hat{\sigma}_j^z \rangle \langle \hat{\sigma}_{j\pm1}^\pm \rangle$. This approximation truncates the otherwise infinite hierarchy of equations, yielding a closed set of nonlinear differential equations for the expectation values. Crucially, we preserve the nearest-neighbour spin-spin correlators such as $\langle \hat{\sigma}_{j}^\pm \hat{\sigma}_{j\pm1}^\pm \rangle$, which are essential in stimulating population inversions and driving the system into a lasing phase.

To ascertain the universality of synchronization phenomena, we extend our analysis to a large system comprising $N$ spins and $N$ MOs. The dynamics of this system is governed by the following set of self-consistent kinetic equations (together with $6N$ further equations for the conjugate operators). 

\begin{align}
\frac{d\langle \hat{\sigma}_j^+ \rangle}{dt} &= i \left\{ \Delta_j - 2\lambda_j ( \langle \hat{b}_j^\dagger \rangle + \langle \hat{b}_j \rangle ) + \frac{i\Gamma}{2}\right \} \langle \hat{\sigma}_j^+ \rangle - i \langle \hat{\sigma}_j^z \rangle \left\{ J_{j-1} \cos{(\Omega_{j-1} t)} \left( \langle  \hat{\sigma}_{j-1}^+ \rangle + \langle  \hat{\sigma}_{j-1}^- \rangle \right ) + J_j \cos{(\Omega_j t)}   \left( \langle \hat{\sigma}_{j+1}^+ \rangle + \langle  \hat{\sigma}_{j+1}^- \rangle \right ) \right \}, \nonumber\\
\frac{d\langle \hat{\sigma}_j^z \rangle}{dt} &= 2i J_{j-1} \cos{(\Omega_{j-1}t)} \left( \langle \hat{\sigma}_{j-1}^+ \hat{\sigma}_j^- \rangle + \langle \hat{\sigma}_{j-1}^- \hat{\sigma}_j^- \rangle - \langle \hat{\sigma}_{j-1}^+ \hat{\sigma}_j^+ \rangle - \langle \hat{\sigma}_{j-1}^- \hat{\sigma}_j^+ \rangle  \right) \nonumber \\ 
& + 2i J_j \cos{(\Omega_j t)} \left( \langle \hat{\sigma}_j^- \hat{\sigma}_{j+1}^+ \rangle + \langle \hat{\sigma}_j^- \hat{\sigma}_{j+1}^- \rangle - \langle \hat{\sigma}_j^+ \hat{\sigma}_{j+1}^+ \rangle - \langle \hat{\sigma}_j^+ \hat{\sigma}_{j+1}^- \rangle  \right) - \Gamma\left( \langle \hat{\sigma}_j^z \rangle + 1 \right), \nonumber\\
\frac{d\langle \hat{b}_j \rangle}{dt} &= -\left(i\omega_j + \frac{\gamma}{2} \right)\langle \hat{b}_j \rangle + i\lambda_j\langle \hat{\sigma}_j^z \rangle, \nonumber \\
\frac{d\langle \hat{b}_j^\dagger \hat{b}_j \rangle}{dt} &= i\lambda_j\langle \hat{\sigma}_j^z \rangle\left(\langle \hat{b}_j^\dagger \rangle - \langle \hat{b}_j \rangle\right) - \gamma\langle \hat{b}_j^\dagger \hat{b}_j \rangle, \nonumber \\
\frac{d\langle \hat{b}_{j}^\dagger \hat{b}_{j}^\dagger \hat{b}_{j}\hat{b}_{j} \rangle}{dt} &= -2i\lambda_j \langle \hat{\sigma}_j^z \rangle \left(\langle \hat{b}_j^\dagger \hat{b}_j  \hat{b}_j \rangle - \langle \hat{b}_j^\dagger \hat{b}_j^\dagger \hat{b}_j \rangle \right)  - 2\gamma \langle \hat{b}_{j}^\dagger \hat{b}_j^\dagger\hat{b}_{j}\hat{b}_{j} \rangle, \nonumber \\
\frac{d\langle \hat{b}_j^\dagger \hat{b}_j \hat{b}_j \rangle}{dt} &= -\left(i\omega_j + \frac{3}{2}\gamma\right) \langle \hat{b}_j^\dagger \hat{b}_j \hat{b}_j \rangle - i\lambda_j \langle \hat{\sigma}_j^z \rangle \left( \langle \hat{b}_{j}^2 \rangle - 2\langle \hat{b}_j^\dagger \hat{b}_j  \rangle \right), \nonumber \\
\frac{d\langle \hat{b}_{j}^2\rangle}{dt} &= -\left(2i\omega_j + \gamma\right) \langle \hat{b}_{j}^2 \rangle + 2i\lambda_j \langle \hat{\sigma}_j^z \rangle \langle \hat{b}_j \rangle, \nonumber \\
\frac{d\langle \hat{\sigma}_j^+ \hat{\sigma}_{j+1}^+ \rangle}{dt} &= 2i \left\{ \Delta_j - \lambda_j ( \langle \hat{b}_j^\dagger \rangle + \langle \hat{b}_j \rangle ) - \lambda_{j+1} ( \langle \hat{b}_{j+1}^\dagger \rangle + \langle \hat{b}_{j+1} \rangle )\right \} \langle \hat{\sigma}_j^+ \hat{\sigma}_{j+1}^+ \rangle - i J_{j-1} \cos{(\Omega_{j-1} t)} \langle \hat{\sigma}_j^z \rangle \langle \hat{\sigma}_{j+1}^+ \rangle \left( \langle \hat{\sigma}_{j-1}^+ \rangle + \langle \hat{\sigma}_{j-1}^- \rangle \right)  \nonumber\\ 
&- i J_j/2 \cos{(\Omega_j t)} \left ( \langle \hat{\sigma}_j^z \rangle + \langle \hat{\sigma}_{j+1}^z \rangle \right ) - i J_{j+1} \cos{(\Omega_{j+1} t)} \langle \hat{\sigma}_j^+ \rangle \langle \hat{\sigma}_{j+1}^z \rangle \left( \langle \hat{\sigma}_{j+2}^+ \rangle + \langle \hat{\sigma}_{j+2}^- \rangle \right) - \Gamma \langle \hat{\sigma}_j^+ \hat{\sigma}_{j+1}^+ \rangle, \nonumber\\
\frac{d\langle \hat{\sigma}_j^- \hat{\sigma}_{j+1}^+ \rangle}{dt} &= 2i \left\{ \lambda_j ( \langle \hat{b}_j^\dagger \rangle + \langle \hat{b}_j \rangle ) - \lambda_{j+1} ( \langle \hat{b}_{j+1}^\dagger \rangle + \langle \hat{b}_{j+1} \rangle )\right \} \langle \hat{\sigma}_j^- \hat{\sigma}_{j+1}^+ \rangle + i J_{j-1} \cos{(\Omega_{j-1} t)} \langle \hat{\sigma}_j^z \rangle \langle \hat{\sigma}_{j+1}^+ \rangle \left( \langle \hat{\sigma}_{j-1}^+ \rangle + \langle \hat{\sigma}_{j-1}^- \rangle \right)  \nonumber\\ 
&- i J_j/2 \cos{(\Omega_j t)} \left ( \langle \hat{\sigma}_{j+1}^z \rangle - \langle \hat{\sigma}_{j}^z \rangle \right ) - i J_{j+1} \cos{(\Omega_{j+1} t)} \langle \hat{\sigma}_j^- \rangle \langle \hat{\sigma}_{j+1}^z \rangle \left( \langle \hat{\sigma}_{j+2}^+ \rangle + \langle \hat{\sigma}_{j+2}^- \rangle \right) - \Gamma \langle \hat{\sigma}_j^- \hat{\sigma}_{j+1}^+ \rangle, \label{eq_SM_kinetic}
\end{align}
where the last two expressions in Eq.~\eqref{eq_SM_kinetic}, corresponding to $\frac{d\langle \hat{\sigma}_j^- \hat{\sigma}_{j+1}^+ \rangle}{dt}$ and $\frac{d\langle \hat{\sigma}_j^+ \hat{\sigma}_{j+1}^+ \rangle}{dt}$, consider $j$ in the range $j = 1,\dots,N-1$. The nonlinear system of differential equations (with $J_0=J_N\equiv 0$) can be integrated numerically to obtain the phonon-number expectation values and the second-order coherence. The latter is defined as
$g^{(2)}(0)=\frac{\langle \hat b^\dagger \hat b^\dagger \hat b \hat b\rangle}{\langle \hat b^\dagger \hat b\rangle^{2}}$. 

These results may then be compared with the corresponding quantities computed from the master equation using e.g. the effective Hamiltonian \eqref{eff1}. A key strength of the kinetic-equation approach is that it enables the numerical study of a large array of spins and MOs, including the dynamical synchronization in phonon lasers—an effect inaccessible within the time-independent framework of the effective Hamiltonians.

\subsection{Scalable On-Demand Phonon Laser Arrays at Any Frequency}

In this subsection, we provide numerical evidence for phonon lasing in multiple mechanical oscillators (MOs) that are physically coupled to the spin ensemble (i.e., $\lambda>0$). We focus on the case of a doughnut-shaped lasing mode, as detailed in Sec. \hyperref[AppendA]{A} (Case I). For lasing to occur in a given MO mode with $\omega_k$, the resonance condition $\sum_{j}^{j+1}\Delta_{k} + \omega_{k} = \pm \Omega_{j}$, with $k=(j,j+1)$, must be satisfied. This relation offers significant flexibility, as it permits the selection of MOs across a wide range of frequencies. However, this freedom comes at a cost: the lasing modes do not synchronize. In contrast to the phase-locked, coherent displaced-state lasing discussed in the main text, the phases of these individual MOs remain diffused, making any pairwise or collective synchronization impossible.

\begin{figure}[t]
\centering
\includegraphics[width=0.72\linewidth]{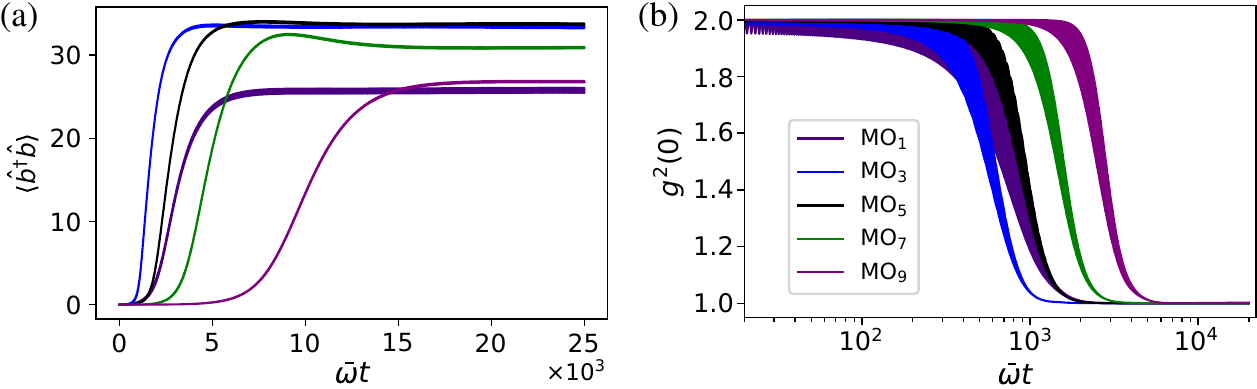}
\caption{(a) Phonon expectation and (b) second order correlation function vs time for $N=10$ mechanical oscillators. Here, we observe that lasing generation is stimulated for those mechanical oscillators where the resonance condition $\Omega_j = \sum_{j}^{j+1}\Delta_{k} + \omega_k$ is fulfilled. Other mechanical oscillators remain decoupled from the spin chains and can be activated on demand. The parameters (in units of $\overline\omega$) are: $\Delta_k=2$, $\omega_j=\{5, 0, 7, 0, 9, 0, 12, 0, 16, 0\}$, $\Omega_j=\{4+\omega_1, 4+\omega_3, 4+\omega_3, 4+\omega_5, 4+\omega_5, 4+\omega_7, 4+\omega_7, 4+\omega_9, 4+\omega_9, 0\}$, $\lambda_j = \{0.4,0,0.4,0,0.4,0,0.4,0,0.4,0\}$, $J_j=0.3$, $\Gamma_{j}=8\times10^{-2}$, $\gamma_j=10^{-3}$.}
\label{figS2}
\end{figure}

Fig. \ref{figS2} presents the key results, showing the temporal evolution of both the average phonon number, $\langle b^\dagger b\rangle$, and the second-order coherence function, $g^{(2)}(0)$. The behaviour of these quantities confirms the establishment of a phonon lasing state. 

\section{Phonon laser using Local Driving on a Single Spin}
This section expands on the \textit{Experimental feasibility} discussion of the main text, where we examine how the elemental phonon laser can be implemented by applying a local drive to a single spin, sketched by the green sphere in Fig. \ref{figS3}\hyperref[figS3]{(a)}. This driven spin interacts with a neighbouring spin through the coupling constant $J_1$, while the latter is coupled to a mechanical oscillator with strength $\lambda_1$. The corresponding Hamiltonian takes the form ($\hbar = 1$):
\begin{align}\label{eq_local_driven_hamiltonian}
    \hat{\mathcal{H}}_L&=\mathcal{\hat{H}}_0+\mathcal{\hat{H}}_0(t)+\mathcal{\hat{H}}_1.       
\end{align}
where $\mathcal{\hat{H}}_0=\sum_{j=1}^2 \frac{\Delta_j}{2} \hat{\sigma}_{j}^{z} + \omega_1\hat{b}_1^{\dagger}\hat{b}_1-\lambda_{1}\hat{\sigma}_{2}^{z}(\hat{b}_1^{\dagger}+\hat{b}_1)$, $\mathcal{\hat{H}}_I=J_1\hat{\sigma}_{1}^{x}\hat{\sigma}_{2}^{x}$, $\mathcal{\hat{H}}_0(t)=\varepsilon_0\cos{(\nu t)} \hat{\sigma}_{1}^{z}$. Here, $\varepsilon_0$ and $\nu$ denote the amplitude and frequency of the local driving field, respectively.

\subsection{Effective Hamiltonian}
We assume identical spin gaps, $\Delta_1 = \Delta_2 = 2\overline\omega$. To derive the effective Hamiltonian, we move to the rotating frame with respect to $\hat{\mathcal{H}}_0(t)$. Since $[\mathcal{\hat{H}}_0,\mathcal{\hat{H}}_0(t)]=0$, and the Hamiltonian $\mathcal{\hat{H}}_0(t)$ commutes with itself at different times $t_1\ne t_2$, that is, $[\mathcal{\hat{H}}_0(t_1),\mathcal{\hat{H}}_0(t_2)]=0$, the unitary transformation reads $\mathcal{\hat{U}}(t)=e^{-\frac{i}{\hbar}\int_{0}^{t}\mathcal{\hat{H}}_0(s)ds}$. This leads us to the Hamiltonian in the rotating frame
\begin{eqnarray}
\hat{\mathcal{H}}_{rot,1}&=&\mathcal{\hat{H}}_0+J_1(e^{2i\mathcal{F}(t)}\sigma^{+}_1+e^{-2i\mathcal{F}(t)}\sigma^{-}_1)(\sigma_2^{+}+\sigma_2^{-}),
\end{eqnarray}
where $\mathcal{F}(t)=(\varepsilon_0/\nu)\sin{\nu t}$. Now, we make use of the Jacobi-Anger expansion $e^{iz\sin{\theta}}=\sum_{m\in\mathbf{Z}}\mathcal{J}_m(z)e^{im\theta}$, where $\mathcal{J}_m(z)$ are first-order Bessel functions, and rewrite the Hamiltonian as follows
\begin{equation}
    \mathcal{\hat{H}}_{{\rm rot},1}(t)= \mathcal{\hat{H}}_0+J_1\bigg(\sum_{m\in\mathbf{Z}}\mathcal{J}_m(2\varepsilon_0/\nu)e^{i m\nu t}\sigma^{+}_1+\sum_{m\in\mathbf{Z}}\mathcal{J}_m(2\varepsilon_0/\nu)e^{-i m\nu t}\sigma^{-}_1\bigg)(\sigma^{+}_2+\sigma^{-}_2).
\end{equation}
Now, we move to the second interaction picture: $\hat{\mathcal{H}}_{rot,2}=e^{i \hat{\mathcal{H}}'_{0} t}\hat{\mathcal{H}}'_{1}e^{-i \hat{\mathcal{H}}'_{0} t}$, where 
\begin{eqnarray}
\hat{\mathcal{H}}'_{0}&=&\overline\omega\sum_{j=1}^2 \hat{\sigma}_j^z 
    + \omega_1\hat{b}_1^\dagger \hat{b}_1, \\ \hat{\mathcal{H}}'_{1}&=&J_1\bigg(\sum_{m\in\mathbf{Z}}\mathcal{J}_m(2\varepsilon_0/\nu)e^{i m\nu t}\sigma^{+}_1+\sum_{m\in\mathbf{Z}}\mathcal{J}_m(2\varepsilon_0/\nu)e^{-i m\nu t}\sigma^{-}_1\bigg)(\sigma^{+}_2+\sigma^{-}_2)-\lambda_{1}\hat{\sigma}_{2}^{z}(\hat{b}_1^{\dagger}+\hat{b}_1).
\end{eqnarray}

With the above Hamiltonians and interaction picture transformation, one gets: $\hat{\mathcal{H}}_{rot,2}=\hat{\mathcal{V}}_{0}+\hat{\mathcal{V}}_{1}$, where
\begin{align}
    \hat{\mathcal{V}}_{0}&=-i \lambda_1\,\hat{\sigma}_{2}^{z}\,\left(\hat{b}_1^{\dagger}\,e^{i\omega_1 t}-\hat{b}_1\,e^{-i\omega_1 t}\right)\equiv\hat{\sigma}_{2}^{z}\,\hat{f}_1(t),\\
    \hat{\mathcal{V}}_{1}&=J_1\bigg(\sum_{m\in\mathbf{Z}}\mathcal{J}_m(2\varepsilon_0/\nu)e^{i m\nu t}e^{4i\overline\omega t}\sigma^{+}_1\sigma^{+}_2+\sum_{m\in\mathbf{Z}}\mathcal{J}_m(2\varepsilon_0/\nu)e^{i m\nu t}\sigma^{+}_1\sigma^{-}_2 + {\rm H.c.}\bigg),
\end{align}
with $\hat{f}_1(t)=-i\lambda_1\,\left(\hat{b}_1^{\dagger}\,e^{i\omega_1t}-\hat{b}_1\,e^{-i\omega_1t}\right)$.

Next, we move to a third interaction picture, which is defined as follows:
\begin{eqnarray}\label{2ndnew}    \hat{\mathcal{V}}'&=&\exp{\left\{i\int\hat{\mathcal{V}}_{0}dt\right\}}\hat{\mathcal{V}}_{1}\exp{\left\{-i\int\hat{\mathcal{V}}_{0}dt\right\}}\nonumber\\
&=&J_1\Big(\sum_{m\in\mathbf{Z}}\mathcal{J}_m(2\varepsilon_0/\nu)e^{i m\nu t}e^{4i\overline\omega t}e^{i\hat{\sigma}_{2}^{z}\hat{F}_1(t)}\hat{\sigma}_1^{+} \hat{\sigma}_2^{+}e^{-i\hat{\sigma}_{2}^{z}\hat{F}_1(t)}+\sum_{m\in\mathbf{Z}}\mathcal{J}_m(2\varepsilon_0/\nu)e^{i m\nu t}e^{i\hat{\sigma}_{2}^{z}\hat{F}_1(t)}\hat{\sigma}_1^{+} \hat{\sigma}_2^{-}e^{-i\hat{\sigma}_{2}^{z}\hat{F}_1(t)}\Big)+ H.c. \nonumber\\
&\equiv& \hat{\mathcal{V}}'_{I}+\hat{\mathcal{V}}'_{II} + H.c.,
\end{eqnarray}
with the Hermitian operator $\hat{F}_1(t)=\frac{\lambda_1}{\omega_{1}}\left(\hat{b}_1^{\dagger}\eta_1+\hat{b}_1\eta_1^{*}\right)$, where $\eta_1=e^{i\omega_{1}t}-1$. In deriving the effective Hamiltonian, we assume that the spin–mechanical coupling $\lambda_1$ is much weaker than the mechanical frequency $\omega_1$, enabling the following analytical simplification: $e^{i \hat{F}_1(t)}\approx1+i\frac{\lambda_1}{\omega_{1}}\left(\hat{b}_{1}^{\dagger}\eta_{1}+\hat{b}_{1}\eta_{1}^{*}\right)$. 

In what follows, the explicit form of each operator $\hat{\mathcal{V}}'$ appearing in Eq.~\eqref{2ndnew} is derived:
\begin{eqnarray}
    \hat{\mathcal{V}}'_{I}&\equiv& J_1\sum_{m\in\mathbf{Z}}\mathcal{J}_m(2\varepsilon_0/\nu)e^{i m\nu t}e^{4i \overline\omega t}e^{i\hat{\sigma}_{2}^{z}\hat{F}_1(t)}\hat{\sigma}_1^{+} \hat{\sigma}_2^{+}e^{-i\hat{\sigma}_{2}^{z}\hat{F}_1(t)}=J_1\sum_{m\in\mathbf{Z}}\mathcal{J}_m(2\varepsilon_0/\nu)e^{i m\nu t}e^{4i \overline\omega t}\hat{\sigma}_1^{+} \hat{\sigma}_2^{+}e^{2i\hat{F}_1(t)}\nonumber\\
    &=& J_1\sum_{m\in\mathbf{Z}}\mathcal{J}_m(2\varepsilon_0/\nu)e^{i m\nu t}e^{4i \overline\omega t}\hat{\sigma}_1^{+} \hat{\sigma}_2^{+}\Big[1+2i\frac{\lambda_1}{\omega_{1}}\left(\hat{b}_1^{\dagger}\eta_1+\hat{b}_1\eta_1^{*}\right)\Big]\nonumber\\
    &=& J_1\sum_{m\in\mathbf{Z}}\mathcal{J}_m(2\varepsilon_0/\nu)e^{i m\nu t}e^{4i \overline\omega t}\hat{\sigma}_1^{+} \hat{\sigma}_2^{+}\nonumber\\
    &+& 2i\frac{J_{1}\lambda_1}{\omega_{1}}\sum_{m\in\mathbf{Z}}\mathcal{J}_m(2\varepsilon_0/\nu)\Big[\hat{\sigma}_1^{+}\hat{\sigma}_2^{+}\hat{b}_1^{\dagger}e^{i(m\nu+4\overline\omega+\omega_{1})t}-\hat{\sigma}_1^{+}\hat{\sigma}_2^{+}\left(\hat{b}_1^{\dagger}+\hat{b}_1\right)e^{i(m\nu+4\overline\omega)t}+\hat{\sigma}_1^{+}\hat{\sigma}_2^{+}\hat{b}_1e^{i(m\nu+4\overline\omega-\omega_{1})t}\Big],
\end{eqnarray}

\begin{eqnarray}
    \hat{\mathcal{V}}'_{II}&\equiv&  J_1\sum_{m\in\mathbf{Z}}\mathcal{J}_m(2\varepsilon_0/\nu)e^{i m\nu t}e^{i\hat{\sigma}_{2}^{z}\hat{F}_1(t)}\hat{\sigma}_1^{+} \hat{\sigma}_2^{-}e^{-i\hat{\sigma}_{2}^{z}\hat{F}_1(t)}=J_1\sum_{m\in\mathbf{Z}}\mathcal{J}_m(2\varepsilon_0/\nu)e^{i m\nu t}e^{4i \overline\omega t}\hat{\sigma}_1^{+} \hat{\sigma}_2^{-}e^{-2i\hat{F}_1(t)}\nonumber\\
    &=& J_1\sum_{m\in\mathbf{Z}}\mathcal{J}_m(2\varepsilon_0/\nu)e^{i m\nu t}\hat{\sigma}_1^{+} \hat{\sigma}_2^{-}\Big[1-2i\frac{\lambda_1}{\omega_{1}}\left(\hat{b}_1^{\dagger}\eta_1+\hat{b}_1\eta_1^{*}\right)\Big]\nonumber\\
    &=& J_1\sum_{m\in\mathbf{Z}}\mathcal{J}_m(2\varepsilon_0/\nu)e^{i m\nu t}\hat{\sigma}_1^{+} \hat{\sigma}_2^{-}\nonumber\\
    &-& 2i\frac{J_{1}\lambda_1}{\omega_{1}}\sum_{m\in\mathbf{Z}}\mathcal{J}_m(2\varepsilon_0/\nu)\Big[\hat{\sigma}_1^{+}\hat{\sigma}_2^{-}\hat{b}_1^{\dagger}e^{i(m\nu+\omega_{1})t}-\hat{\sigma}_1^{+}\hat{\sigma}_2^{-}\left(\hat{b}_1^{\dagger}+\hat{b}_1\right)e^{im\nu t}+\hat{\sigma}_1^{+}\hat{\sigma}_2^{-}\hat{b}_1e^{i(m\nu-\omega_{1})t}\Big].
\end{eqnarray}

In the following, we apply the Floquet theory to time-periodic Hamiltonians \cite{Bukov2015AdvPhys, Floquet_1883}. The main features of the
system dynamics can be captured by the one-period evolution operator $\hat{U}(T)=e^{-i\hat{H}_{F}T/\hbar}$. where $\hat{H}_{F}$ is the time-independent Floquet Hamiltonian. In the high-frequency regime, $\hat{H}_{F}$ can be approximated using the Magnus expansion $\hat{H}_{F}=\sum_{l=0}^{\infty}\hat{H}_{F}^{(l)}$. The first term read as:
\begin{eqnarray*}
    \hat{H}_{F}^{(0)}&=&\frac{1}{T}\int_0^T dt \hat{\mathcal{V}}'(t)\nonumber\\    &\simeq&J_1\bigg(\sum_{m\in\mathbf{Z}}\mathcal{J}_m(2\varepsilon_0/\nu)\delta_{m\nu+4\overline\omega,0}\hat{\sigma}_1^{+} \hat{\sigma}_2^{+}+\sum_{m\in\mathbf{Z}}\mathcal{J}_m(2\varepsilon_0/\nu)\delta_{m\nu,0}\hat{\sigma}_1^{+} \hat{\sigma}_2^{-}\bigg)\nonumber\\
    &+& 2i\frac{J_{1}\lambda_1}{\omega_{1}}\bigg(\sum_{m\in\mathbf{Z}}\mathcal{J}_m(2\varepsilon_0/\nu)\delta_{m\nu+4\overline\omega+\omega_1,0}\hat{\sigma}_1^{+}\hat{\sigma}_2^{+}\hat{b}_1^{\dagger}-\sum_{m\in\mathbf{Z}}\mathcal{J}_m(2\varepsilon_0/\nu)\delta_{m\nu+4\overline\omega,0}\hat{\sigma}_1^{+}\hat{\sigma}_2^{+}\left(\hat{b}_1^{\dagger}+\hat{b}_1\right)+\sum_{m\in\mathbf{Z}}\mathcal{J}_m(2\varepsilon_0/\nu)\delta_{m\nu+4\overline\omega-\omega_1,0}\hat{\sigma}_1^{+}\hat{\sigma}_2^{+}\hat{b}_1\bigg)\\
    &-& 2i\frac{J_{1}\lambda_1}{\omega_{1}}\bigg(\sum_{m\in\mathbf{Z}}\mathcal{J}_m(2\varepsilon_0/\nu)\delta_{m\nu+\omega_1,0}\hat{\sigma}_1^{+}\hat{\sigma}_2^{-}\hat{b}_1^{\dagger}-\sum_{m\in\mathbf{Z}}\mathcal{J}_m(2\varepsilon_0/\nu)\delta_{m\nu,0}\hat{\sigma}_1^{+}\hat{\sigma}_2^{-}\left(\hat{b}_1^{\dagger}+\hat{b}_1\right)+\sum_{m\in\mathbf{Z}}\mathcal{J}_m(2\varepsilon_0/\nu)\delta_{m\nu-\omega_1,0}\hat{\sigma}_1^{+}\hat{\sigma}_2^{-}\hat{b}_1\bigg) + H.c.\\
    &=&J_1\bigg(\mathcal{J}_{\alpha}(2\varepsilon_0/\nu)\hat{\sigma}_1^{+} \hat{\sigma}_2^{+}+\mathcal{J}_0(2\varepsilon_0/\nu)\hat{\sigma}_1^{+} \hat{\sigma}_2^{-}\bigg)\nonumber\\
    &+& 2i\frac{J_{1}\lambda_1}{\omega_{1}}\bigg(\mathcal{J}_{\beta}(2\varepsilon_0/\nu)\hat{\sigma}_1^{+}\hat{\sigma}_2^{+}\hat{b}_1^{\dagger}-\mathcal{J}_{\alpha}(2\varepsilon_0/\nu)\hat{\sigma}_1^{+}\hat{\sigma}_2^{+}\left(\hat{b}_1^{\dagger}+\hat{b}_1\right)+\mathcal{J}_{\delta}(2\varepsilon_0/\nu)\hat{\sigma}_1^{+}\hat{\sigma}_2^{+}\hat{b}_1\bigg)\\
    &-& 2i\frac{J_{1}\lambda_1}{\omega_{1}}\bigg(\mathcal{J}_{\eta}(2\varepsilon_0/\nu)\hat{\sigma}_1^{+}\hat{\sigma}_2^{-}\hat{b}_1^{\dagger}-\mathcal{J}_0(2\varepsilon_0/\nu)\hat{\sigma}_1^{+}\hat{\sigma}_2^{-}\left(\hat{b}_1^{\dagger}+\hat{b}_1\right)+\mathcal{J}_{\kappa}(2\varepsilon_0/\nu)\hat{\sigma}_1^{+}\hat{\sigma}_2^{-}\hat{b}_1\bigg)+ H.c.
\end{eqnarray*}
with $\mathbb{Z}$ number: $\alpha=-4\overline\omega/\nu$, $\beta=-(4\overline\omega+\omega_1)/\nu$, $\delta=-(4\overline\omega-\omega_1)/\nu$, $\eta=-\omega_1/\nu$ and $\kappa=\omega_1/\nu$.

As discussed in Sec. \hyperref[AppendA2]{A.2.}, carefully enforcing well-defined resonance conditions allows for the selective realization of two distinct lasing regimes. These regimes are characterized by (I) a displaced $\alpha$-coherent state emission profile and (II) a doughnut‑shaped mode exhibiting an indeterminate phase distribution.

\textit{Case I.} By taking $\varepsilon_{0} = (\nu/2)x_{0}$ with $x_{0}=2.4048$, the first root of the zeroth-order Bessel function $\mathcal{J}_{0}(x)$, and using $\omega_{1}=8\overline\omega$ and $\nu=12\overline\omega$, we obtain $\beta=-1$ (note that $\left\{\alpha,\delta,\eta,\kappa\right\} \notin \mathbb{Z}$). Under these conditions, the resulting effective Hamiltonian becomes equivalent to the doughnut-shaped phonon-mode laser $\mathcal{\hat{H}}_{I}^{eff}$ in Eq. (\ref{effdona}):
\begin{equation} \label{eq:heff2_l}
    \hat{H}_{F,I}^{(0)}\simeq 2i\frac{J_1\lambda_1}{\omega_1}\mathcal{J}_{-1}(2\varepsilon_0/\nu)\hat{\sigma}_{1}^{+}\hat{\sigma}_{2}^{+}\hat{b}_1^{\dagger} + H.c.
\end{equation}

\textit{Case II.} By taking $\varepsilon_{0} = (\nu/2)x_{0}$, where $x_{0}=2.4048$ is the first root of the zeroth-order Bessel function $\mathcal{J}_{0}(x)$, and using the frequencies $\omega_{1}=8\overline\omega$ and $\nu=2\overline\omega$, we obtain the parameters $\alpha=-2$, $\beta=-6$, $\delta=2$, $\eta=-4$, and $\kappa=4$ (note that $2\mathcal{J}_2(2\varepsilon_0/\nu)\sim 1$ while $\left\{2\mathcal{J}_{\pm4}(2\varepsilon_0/\nu), 2\mathcal{J}_6(2\varepsilon_0/\nu)\right\}\ll1$ ). Under these conditions, we obtain the effective Hamiltonian governing the phonon phase-locked coherent state, which is equivalent to $\mathcal{\hat{H}}_{II}^{eff}$ in Eq. (\ref{effcoh}):
\begin{equation} \label{eq:heff1_l}
    \hat{H}_{F,II}^{(0)}\simeq J_1\mathcal{J}_2(2\varepsilon_0/\nu)\hat{\sigma}_{1}^{+}\hat{\sigma}_{2}^{+}-2i\frac{J_1\lambda_1}{\omega_1}\mathcal{J}_2(2\varepsilon_0/\nu)\hat{\sigma}_{1}^{+}\hat{\sigma}_{2}^{+}\hat{b}_1^{\dagger}+H.c.
\end{equation}

\subsection{Numerical results} 
Finally, to evaluate the signatures of phonon lasing, we include all relevant dissipation channels and numerically solve the ME:
\begin{eqnarray}\label{eq:master_equation}
\frac{d\hat{\rho}}{dt}&=&-i[\hat{\mathcal{H}}_L,\hat{\rho}]+\sum_{j=1}^2\Gamma_{j}\Bigg[\left(1+\bar{n}_{j}^{s}\right)\mathcal{L}_{\hat{\sigma}_{j}^{-}}[\hat{\rho}]+\bar{n}_{j}^{s}\,\mathcal{L}_{\hat{\sigma}_{j}^{+}}[\hat{\rho}]\Bigg]
+\gamma_{1}\Bigg[\left(1+\bar{n}_{1}^{m}\right)\mathcal{L}_{\hat{b}_1}[\hat{\rho}]+\bar{n}_{1}^{m}\,\mathcal{L}^\dagger_{\hat{b}_1}[\hat{\rho}]\Bigg],
\end{eqnarray}
where the Hamiltonian, $\hat{\mathcal{H}}_L$, is defined in Eq.~\eqref{eq_local_driven_hamiltonian}. $\Gamma_{j}$ and $\gamma_1$ are the spin decay and mechanical damping rates, respectively;  $\bar{n}_{j}^{s}$ ($\bar{n}_{1}^{m}$) specifies the mean thermal occupation of the spins (mechanical) reservoir. 
 
\begin{figure*}[t]
\centering
\includegraphics[width=1\linewidth]{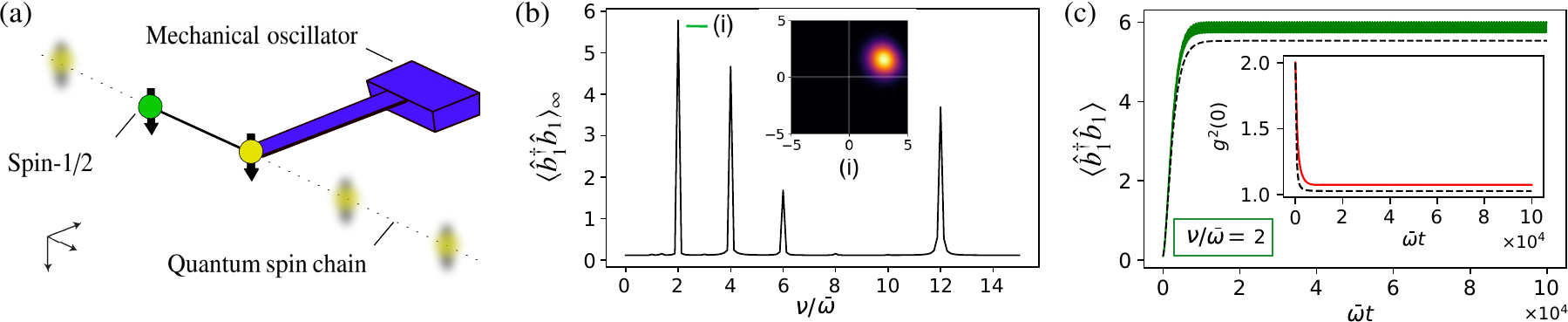}
\caption{(a) Configuration where the first spin is locally driven by $\varepsilon(t)$, indicated by the green sphere. The coupling between spins is assumed to be constant, and the second spin is coupled to the mechanical oscillator. $b)$ Steady-state excitation number of mechanical oscillator as a function of $\nu$. Here $\varepsilon_0=(\nu/2)x_0$ with $x_0=2.4048$. $c)$ Dynamics of excitations and (inset) second order correlation function. The dashed curves correspond to results obtained from the effective Hamiltonian, $\hat{H}_{F,I}^{(0)}$. The parameters (in units of $\overline\omega$) are: $\omega_1=8$, $\lambda_1=0.4$, $J_1=0.12$, $\nu=2$, $\Gamma_{1}=\Gamma_{2}=8\times10^{-2}$, $\gamma_1=10^{-3}$, $\bar{n}_{1}^{s}=\bar{n}_{2}^{s}=0.01$ and $\bar{n}_{1}^{m}=0.1$.}
\label{figS3}
\end{figure*}

In Fig.~\ref{figS3}\hyperref[figS3]{(b)} we evaluate steady-state excitation number of mechanical oscillator as a function of $\nu$. Here, we identify that mechanical gain emerges, for instance, when $\nu/\overline\omega = \{2, 12\}$, as predicted by effective Hamiltonians. The Wigner function (see inset) corresponds to the condition $\nu=2\overline\omega$, consistent with the phase-locked coherent state (II) described by Eq. \eqref{eq:heff1_l}. In Fig. \ref{figS3}\hyperref[figS3]{(c)}, we evaluate the dynamics of excitations and the second-order correlation function of MO, both of which serve as signatures confirming that phonon lasing is achieved through local driving. Additional peaks at $\nu/\overline\omega = \{4, 6\}$ reveal further regimes in which amplification is observed, although these cases are not analysed here. 

\end{document}